\documentclass[a4paper,fleqn,usenatbib]{mnras}

\usepackage[T1]{fontenc}
\usepackage{ae,aecompl}
\usepackage{graphicx}
\usepackage{amsmath}
\usepackage{tabularx}
\usepackage{inputenc}
\title[Characterisation of the radial velocity signal induced by rotation in late-type dwarfs]{Characterisation of the radial velocity signal induced by rotation in late-type dwarfs}

\author[A. Su\'{a}rez Mascare\~{n}o]{A. Su\'{a}rez Mascare\~{n}o$^{1,2,3}$ \thanks{E-mail: Alejandro.SuarezMascareno@unige.ch}, R. Rebolo$^{1,2,4}$, J.~I. Gonz\'alez Hern\'andez$^{1,2}$, M. Esposito$^{4}$
\\
$^{1}$ Instituto de Astrof\'{i}sica de Canarias, E-38205 La Laguna, Tenerife, Spain\\
$^{2}$ Universidad de La Laguna, Dpto. Astrof\'{i}sica, E-38206 La Laguna, Tenerife, Spain\\
$^{3}$ Observatoire Astronomique de l'Universit\'{e} de Gen\'{e}ve, Versoix, Switzerland  \\
$^{4}$ Consejo Superior de Investigaciones Cient{\'\i}ficas, Spain\\
$^{5}$ INAF - Osservatorio Astronomico di Capodimonte, Via Moiariello, 16, 80131 - NAPOLI 
}

\begin{document}

\date{Revised 02/2017}

\pagerange{\pageref{firstpage}--\pageref{lastpage}} \pubyear{2002}

\maketitle

\label{firstpage}

\begin{abstract}
We investigate the activity induced signals related to rotation in late type stars (FGKM). We analyse the Ca II H\&K, the H$_{\alpha}$ and the radial velocity time-series of 55 stars using the spectra from the HARPS public database and the light-curves provided by the ASAS survey. We search for short term periodic signals in the time-series of activity indicators as well as in the photometric light-curves. Radial velocity data sets are then analysed to determine the presence of activity induced signals. We measure a radial velocity signal induced by rotational modulation of stellar surface features in 37 stars, from late F-type to mid M-type stars. We report an empirical relationship, with some degree of spectral type dependency, between the mean level of chromospheric emission measured by the $\log_{10}(R'_\textrm{HK})$ and the measured radial velocity semi amplitude.  We also report a relationship betweeen the semi amplitude of the chromospheric measured signal and the semi amplitude of the radial velocity induced signal, which strongly depends on the spectral type. We find that for a given strength of chromospheric activity (i.e. a given rotation period) M-type stars tend to induce larger rotation related radial velocity signals than G and K-type stars.

\end{abstract}


\section{Introduction}

High precision radial velocity (RV) measurements give astronomers the possibility of detecting small exoplanets, down to the mass of the Earth. Unfortunately, intrinsic variations  of the magnetic regions on the stellar surface induce radial velocity variations which, if stable over a few rotation periods, can mimic a planetary signal. Recognising and characterising them is key to disentangle the true planet induced signals \citep{Queloz2001,  Dumusque2012, Santos2014, Robertson2014}. 

Activity induced signals depend mainly on the activity level and spectral type of a star. Highly active stars usually rotate faster, inducing shorter period signals with larger amplitudes. These amplitudes also depend on the strength of the magnetic field, which is a function of the spectral type as it depends strongly on the depth of the convective zone. These signals are produced by a variety of physical phenomena in the stellar surface. Among them, spots are particularly relevant for rapidly rotating late-type dwarfs \citep{Saar1997, Noyes1984, Santos2000, Hatzes2002, Desort2007}. For slowly rotating stars plages become another important, but not so well understood, source of variability in the timescales of rotation. RV signals induced by plages and spots are not expected to be well correlated \citep{Meunier2010}.

The amplitude of the so-called radial velocity jitter has been subject of study for many years \citep{Saar1998, Santos2000, Paulson2002, Wright2005, MartinezArnaiz2010, Isaacson2010}. Taking advantage of the precision achieved by HARPS (better than 1 m s$^{-1}$) and the sampling rate of planet hunting surveys, we present here a detailed analysis of this jitter measuring its period and amplitude in RV time-series data of a sample of late-type dwarfs. We study these rotationally-induded periodic signals, with particular attention at the harmonics of the stellar rotation period \citep[see also][]{Boisse2011}, and to the relationships between the measured amplitudes and the level of chromospheric emission ($\log_{10}(R'_\textrm{HK})$). We also investigate the relation between the amplitude of the induced RV signals and the amplitude of the modulation of the chromospheric emission measured as the Mount Wilson S Index \citep{Noyes1984}.

\section[]{Stellar Sample and data}

\begin{figure}
\includegraphics[width=\linewidth]{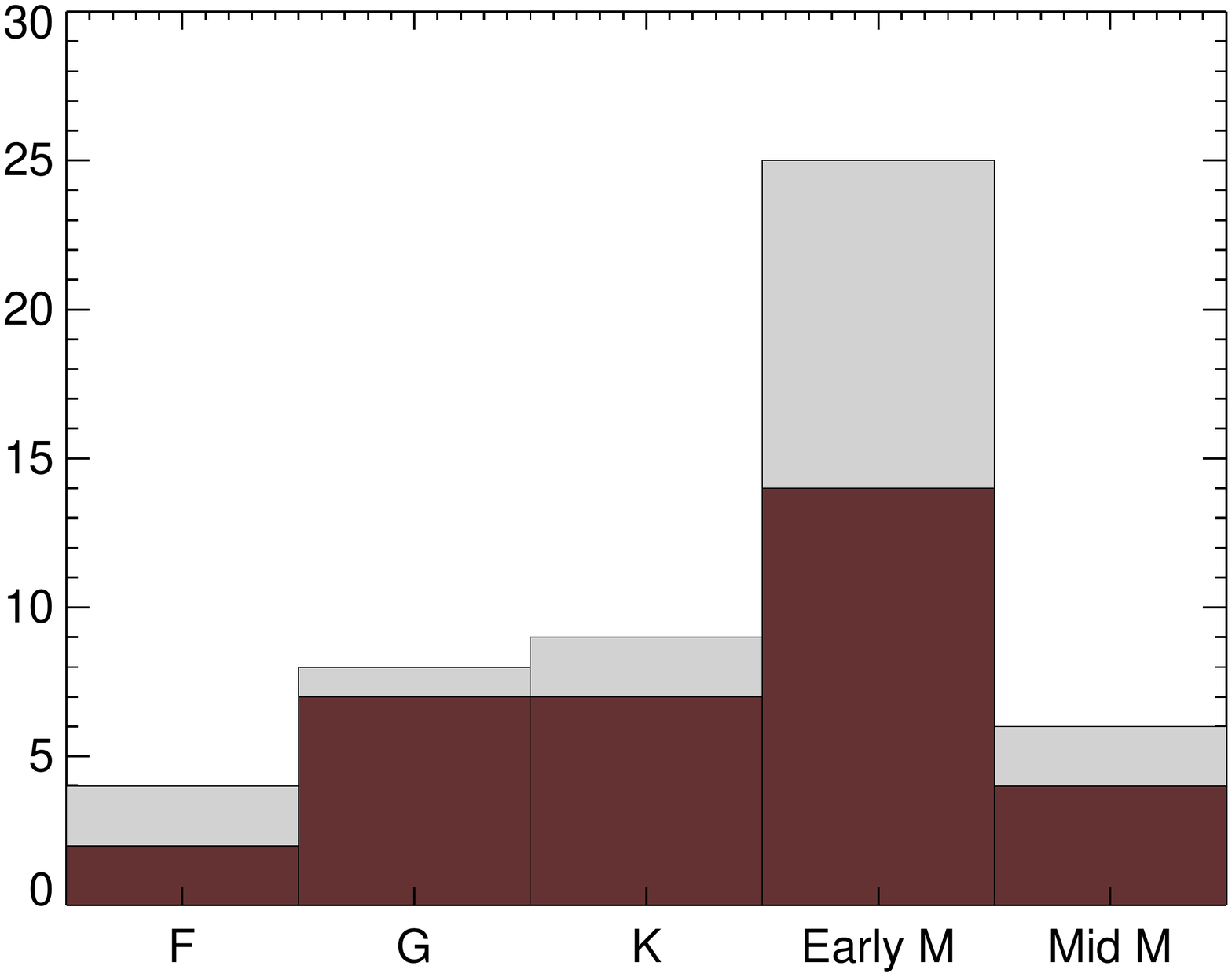}
\caption{Distribution of spectral types in our sample. M-dwarfs have been separated in two groups, with the fully-convective ones (later than M3.5) in a separated category. Dark fill shows the number of detections of periodic RV signals induced by rotation.}
\label{sp_dis}
\end{figure}

We selected a sample of bright southern stars with HARPS spectra available in the ESO public database. Our initial sample consists of 55 low activity ($\log_{10}(R'_\textrm{HK})~ \textless -4.4$) main-sequence stars covering from late-F to mid-M type. The selected stars are part of the planet-hunting programs using HARPS~\citep{Mayor2003} and their rotation periods and magnetic cycles have been investigated using spectroscopic and photometric time-series \citep{Masca2015,Masca2016}. All of them had been observed more than 20 individual nights as for May 2016, giving a total number of more than 9000 spectra. Figure ~\ref{sp_dis} shows the distribution of spectral types for the sample of stars. Table ~\ref{data_sample} shows relevant data for  stars where a reliable detection of an activity induced radial velocity signal was obtained in this work 

\begin {table*}
\begin{center}
\caption {Relevant data for stars in our sample with activity induced radial velocity signals}
    \begin{tabular}{ l  l c c c c c c  l } \hline
Star & Sp. Type  & $m_{B}$ & $m_{V}$  & $N_{Spec}$ & Time Span & $N_{Phot}$ & Time Span &  Ref\\ 
	& 	 &  &  & &  (Years) &&  (Years) & \\\hline
HD 25171		& F8	 	& 8.30	& 7.78	& 50		& 11.3 	& 485 	& 8.8  & 1, 9\\
HD 1581		& F9.5	& 4.80	& 4.23	& 1537	& 9.9 	& 505	& 9.0 & 2, 10\\ \\
HD 1388 		& G0  	& 7.09	& 6.50	& 106	& 9.3	& 335 	& 7.4 & 3, 10	\\
HD 134060 	& G0	 	& 6.91	& 6.29	& 294	& 8.9	& 487 	& 8.6 & 	1, 10 \\
HD 30495		& G1.5	& 6.14	& 5.50	& 82		& 2.4	& 558	& 9.0 & 1, 10	\\
HD 2071		& G2		& 7.97	& 7.27	& 51		& 10.1	& 671	& 8.4 & 1, 10	\\
HD 41248		& G2		& 9.42	& 8.81	& 207	& 10.1	& 1207	& 9.0 & 1, 9	\\
HD 1461		& G3		& 7.14	& 6.46	& 441	& 10.8	& 349	& 8.1 & 	3, 11 \\ 
HD 63765		& G9		& 8.85	& 8.10	& 51		& 6.3	& 604	& 9.0 & 4, 10 \\ \\
Corot-7		& K0		& 12.78	& 11.63	& 161	& 3.9 	& 432	& 8.7 & 5, 12		\\
Alpha Cen B & K1    & 2.21   & 1.33  & 8969  &  5.0  & 1179   & 8.5&2, 17 \\
HD 224789	& K1		& 9.12	& 8.24	& 30		& 7.2 	& 505	& 9.0 & 1, 10\\
HD 40307		& K2.5	& 8.10	& 7.15	& 365	& 10.3	& 871	& 8.9	&  6, 10\\
HD 176986	& K2.5	& 9.39	& 8.45	& 134	& 9.8 & 425 & 8.7	& 1, 10\\
HD 215152	& K3		& 9.12	& 8.13	& 270	& 10.4 & 338 	& 9.0 & 1, 11\\
HD 125595	& K4		& 10.13 & 9.03	& 125	& 5.7 & 736 & 8.7 & 2, 10\\
HD 85512		& K6		& 8.83	& 7.65	& 715	& 5.6 & 720	& 8.9 & 6, 10\\ \\
GJ 676A		& M0		& 11.03	& 9.59	& 109	& 6.4 & 803 & 8.7 & 6, 13\\
GJ 229		& M1 	& 9.607	& 8.125	& 101	& 11.0	& 708	& 9.0 & 6, 14\\
GJ 514		& M1		& 10.52	& 9.03	& 107	& 10.0 & 341	& 8.0 & 6, 14\\
GJ 536		& M1		& 11.18	& 9.71	& 128	& 15.3 & 359	& 8.6 & 6, 14\\
GJ 205		& M1.5	& 9.44	& 7.97	& 71		& 2.9 & 641	& 9.0 & 6, 14\\
GJ 667C		& M1.5	& 11.79	& 10.22	& 180	& 7.9 & 69	& 8.6 & 2, 15\\
GJ 880		& M1.5	& 10.14	& 8.64	& 71		& 10.0 & 230	& 6.4 & 6, 14\\
GJ 526		& M1.5	& 9.89	& 8.50	& 29		& 4.9 & 281 & 6.5 & 6, 14\\
GJ 1			& M2		& 10.02	& 8.56	& 42		& 5.7 & 50	& 1.0 & 6, 16\\
GJ 382		& M2		& 10.76	& 9.26	& 31		& 3.3 & 501	& 9.0 & 6, 7, 14 \\
GJ 832		& M2		& 10.18	& 8.67	& 57		& 8.6 & 486	& 8.9 & 6, 14\\
GJ 176		& M2.5	& 11.59	& 9.95	& 63		& 5.9 & 438 & 7.0 & 6, 16\\
GJ 358		& M3		& 12.22	& 10.69	& 33		& 7.3 & 1001	 & 9.0 & 6, 14\\
GJ 479		& M3		& 12.20	& 10.66	& 54		& 2.9 & 534	& 8.7 & 6, 14\\
GJ 674		& M3		& 12.97	& 9.41	& 131	& 9.8 & 544	& 8.7 & 6, 17\\
GJ 273		& M3.5	& 11.44	& 9.87	& 195	& 10.4 & 379	& 8.0 & 6, 18\\
GJ 876		& M4		& 11.75	& 10.19	& 222	& 8.46 & 387	& 9.0 & 8, 14\\
GJ 581 		& M5		& 11.76	& 10.56	& 219	& 8.0 & 687 & 8.7 & 6, 19\\ 
 \hline
\label{data_sample}
\end{tabular}  
\end{center}
\begin{flushleft}

\textbf{References for magnitudes:} 1 - \citet{Hog2000}, 2 - \citet{Ducati2002}, 3 - \citet{Mermilliod1986}, 4 - \citet{Cousins1962}, 5 - \citet{Leger2009}, 6 - \citet{Koen2010}, 7 - \citet{Kiraga2012}, 8 - \citet{Landolt2009} \\
\textbf{References for spectral types:} 9 - \citet{Houk1975}, 10 - \citet{Gray2006}, 11 - \citet{Gray2003}, 12 - \citet{Ehrenreich2011}, 13 - \citet{Upgren1972}, 14 - \citet{Maldonado2015}, 15 - \citet{Geballe2002}, 15 - \citet{Houk1982}, 16 - \citet{vonBraun2014}, 17 - \citet{Torres2006}, 18 - \citet{Lepine2013}, 19 - \citet{Neves2014}
\end{flushleft}
\end {table*}

\subsection{Spectroscopic Data}

HARPS is a fibre-fed high-resolution echelle spectrograph installed at the 3.6 m ESO telescope in La Silla Observatory (Chile). The instrument has a resolving power $R\sim 115\,000$ over a spectral range from 378 to 691 nm. It has been designed to attain extreme long-term RV accuracy. It is contained inside a vacuum vessel to avoid spectral drifts due to changes in temperature and air pressure. HARPS comes with its own pipeline providing extracted and calibrated spectra, as well as RV measurements and other data products such as cross-correlation functions and their bisector profiles.

For the analysis of the spectral indicators we use the extracted order-by-order wavelength-calibrated spectra produced by the HARPS pipeline. In order to minimize atmospheric effects we create a spectral template for each star to correct the order-by-order flux ratios for the individual spectra. We correct each spectrum for the Earth's barycentric radial velocity and the radial velocity of the star using the measurements given by the pipeline. We finally re-bin them into a constant wavelength step. 

\subsection{Photometry}

We also use the light curves provided by the All Sky Automated Survey (ASAS) public database. ASAS \citep{Pojmanski1997} is an all sky survey in the $V$ and $I$ bands running since 1998 at Las Campanas Observatory, Chile.  Best photometric results are achieved for stars with V $\sim$8-14, but this range can be extended  implementing  some quality control on the data. ASAS has produced light-curves for around $10^{7}$ stars at $\delta < 28^{\circ}$. 

The ASAS catalogue supplies ready-to-use light curves with flags indicating the quality of the data. For this analysis we relied  only on  good quality data (grade "A" and "B" in the internal flags). Even after this quality control,  there are still some high dispersion measurements  which cannot  be explained by a regular stellar behaviour. We reject those measurements by de-trending the series and eliminating points deviating more than three times the standard deviation from the median value.

\section{Stellar activity indicators and Radial Velocities}

We compute the Ca II H\&K index, the $\log_{10}R'_{HK}$ and the H$\alpha$ index following \citet{Noyes1984, Lovis2011, Masca2015} and \citet{Masca2016}. We use the indexes to re-measure the rotation period of the selected stars and to be able to compare the measured radial velocity semi-amplitudes with the activity indicators. 

\subsection{Radial velocities}

Radial velocities are taken directly from the measurements of the standard HARPS pipeline, except for M-type stars. The radial velocity measurements in the HARPS standard pipeline is determined by a Gaussian fit of the cross correlation function (CCF) of the spectrum with a digital mask \citep{PepeMayor2000}. In the case of M-dwarfs, due to the huge number of line blends, the cross correlation function is not Gaussian resulting in a less precise RV measurement and in a loss of sensitivity to the changes in the full width half maximum (FWHM) of the CCF.  For this type of stars we opted to do a different modelling of the cross correlation function, using the combination of a second order polynomial with a Gaussian function over a 15  km s$^{-1}$ window centred at the minimum of the CCF following \citet{Masca2017}. The center of the Gaussian function is taken as our radial velocity measurement. This allows us to improve the stability of the measured radial velocities as well as to improve the sensitivity of the FWHM measurements. For high signal-to-noise measurements the extracted value is virtually the same.

\subsection{Quality control of the data}

As the sampling rate of our data is not  well suited for modelling fast events, such as flares, and their effect in the radial velocity is not well understood, we identify and reject points likely affected by flares by searching for abnormally high measurements of the activity indicators and clear distortions in the Balmer series \citep{Reiners2009}.

\section{Analysis}

\subsection{Radial velocity signals induced by stellar rotation}

In order to identify the rotation induced radial velocity signals we first analyse the time-series of the Ca II H\&K and H$\alpha$ activity indicators, the time-series of the variations of the FWHM of the CCF, and the light curve (when available) to determine the rotation period of the star. Some of the rotation periods were previously measured in \citet{Masca2015, Masca2016}. In these cases we measure it again including the last available data. To search for periodic signals in the series we compute the power spectrum using a generalised Lomb-Scargle periodogram \citep{Zechmeister2009} and if there is any significant periodicity we fit the detected signal using the RVLIN package \citep{WrightHoward2012}.

To evaluate the false alarm probability of any peak in the periodogram we follow \citet{Cumming2004} modification over the work by \citet{HorneBaliunas1986} to obtain the spectral density thresholds for a desired false alarm level. This means our false alarm probability is defined as $FAP = 1 - [1 - P (z > z_{0}]^{M}$ where $P (z > z_{0}) = exp (-z_{0})$ is the  probability of $z$ being greater than $z_{0}$, with $z$ the target spectral density, $z_{0}$ the measured spectral density and $M$ the number of independent frequencies. We search for the power values corresponding to 10 $\%$, 1 $\%$ and 0.1$\%$ false alarm probability. 

The mean rotation period of the star is estimated by calculating the weighted average of the detections obtained with different indicators. Errors are the peak-to-peak variation divided by the square root of the number of detections and the global false alarm probability is the combined false alarm probability given by a Fisher's combined probability test  \citep{fisher1925}.

Different activity proxies provide information about different parts of the stellar atmosphere. Photometry gives information mainly about photospheric variability, while the S$_{MW}$ and H$_{\alpha}$ indexes reflect the variability at different heights of the stellar chromosphere and sometimes even at different latitudes. Thus differential rotation may lead to differences in the rotation period estimates from these three variability indicators which could be larger than the error bars of individual measurements. 
The period of the induced RV signal should be similar but not necessarily coincident with the average period determined from activity proxies. 
 
Once we have a measurement of the rotation period of the star we search for periodic signals in the radial velocity time-series following the same procedure. To do that we first subtract long term trends, if present, using linear fits or incomplete orbits fitting sinusoidal models when the change of slope is apparent. Then we compute the power spectrum using a generalised Lomb-Scargle periodogram \citet{Zechmeister2009} and iteratively fit all the detected signals with keplerian models using the RVLIN package \citep{WrightHoward2012} until there are no significant remaining signals. During the process we were able to recover all the published known planetary signals for the analysed stars (using exoplanet.eu as source). The analysis of their signals is beyond the scope of this paper. After cleaning the series from planetary signals we search for periodic signals compatible with the rotation period previously measured. 

\begin{figure}
\includegraphics[width=9cm]{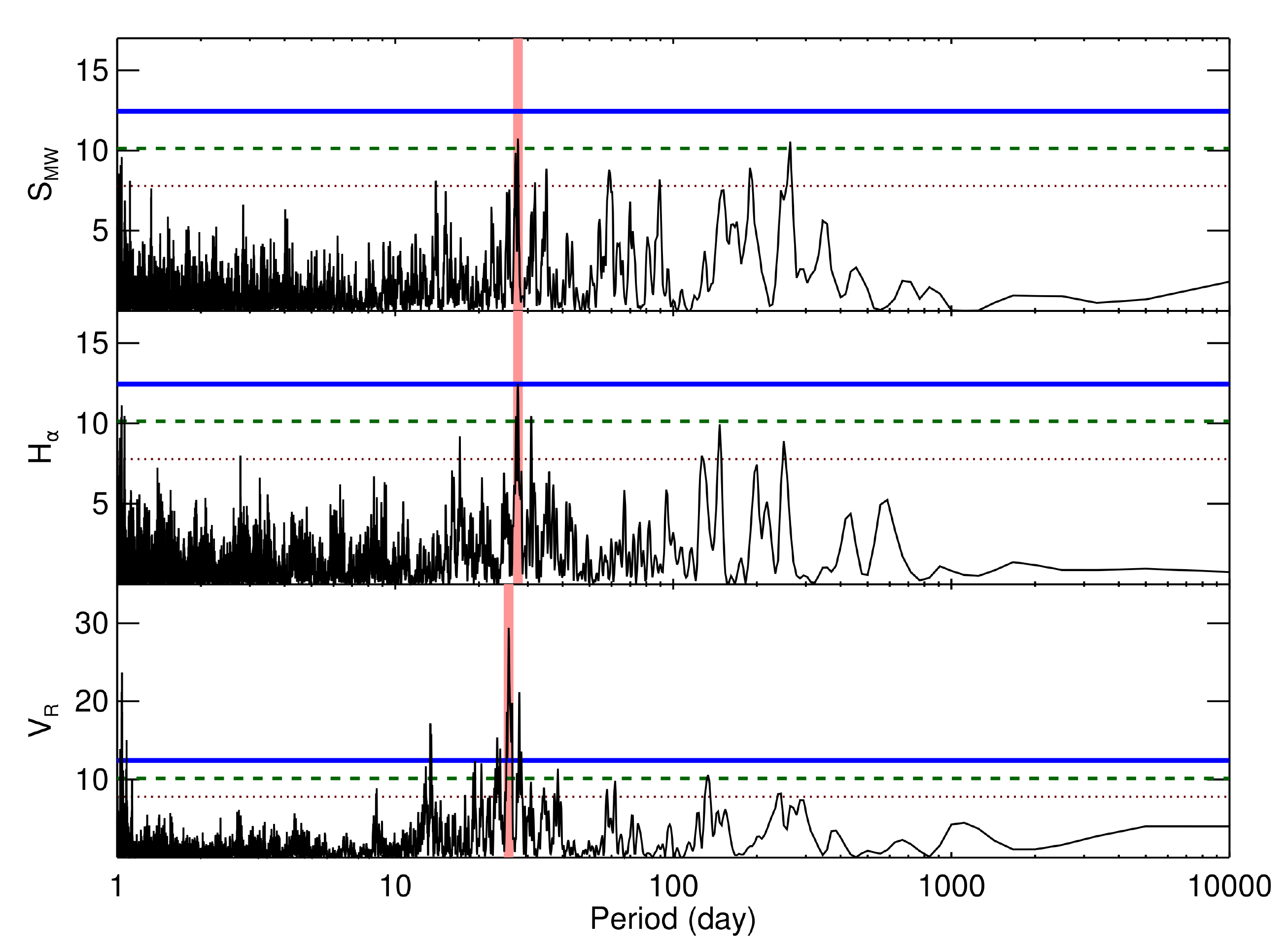}
\caption{Periodograms for the S$_{MW}$ index,  H$_{\alpha}$ index and RV time-series for the G-type star HD 41248. Light red fill identifies the rotational modulation. Horizontal lines show the FAP levels. Red dotted line shows the 10\% FAP level, green dashed line the 1\% level and blue solid line the 0.1\% level.}
\label{HD41248_data}
\end{figure}

\begin{figure}
\includegraphics[width=9cm]{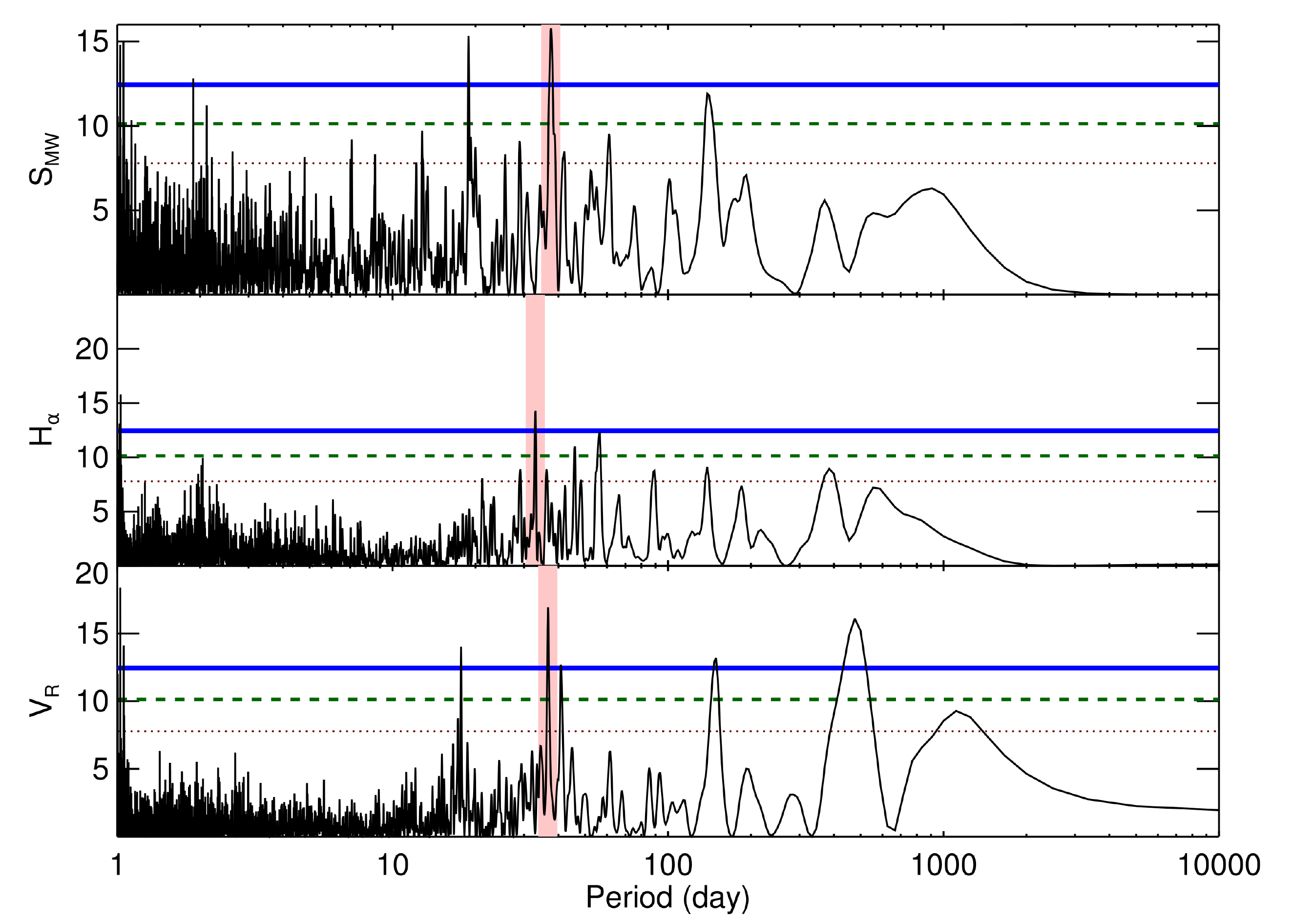}
\caption{Periodograms for the S$_{MW}$ index,  H$_{\alpha}$ index and RV time-series for the K-type star HD 125595. Light red fill identifies the rotational modulation. Horizontal lines show the FAP levels. Red dotted line shows the 10\% FAP level, green dashed line the 1\% level and blue solid line the 0.1\% level.}
\label{HD125595_data}
\end{figure}

\begin{figure}
\includegraphics[width=9cm]{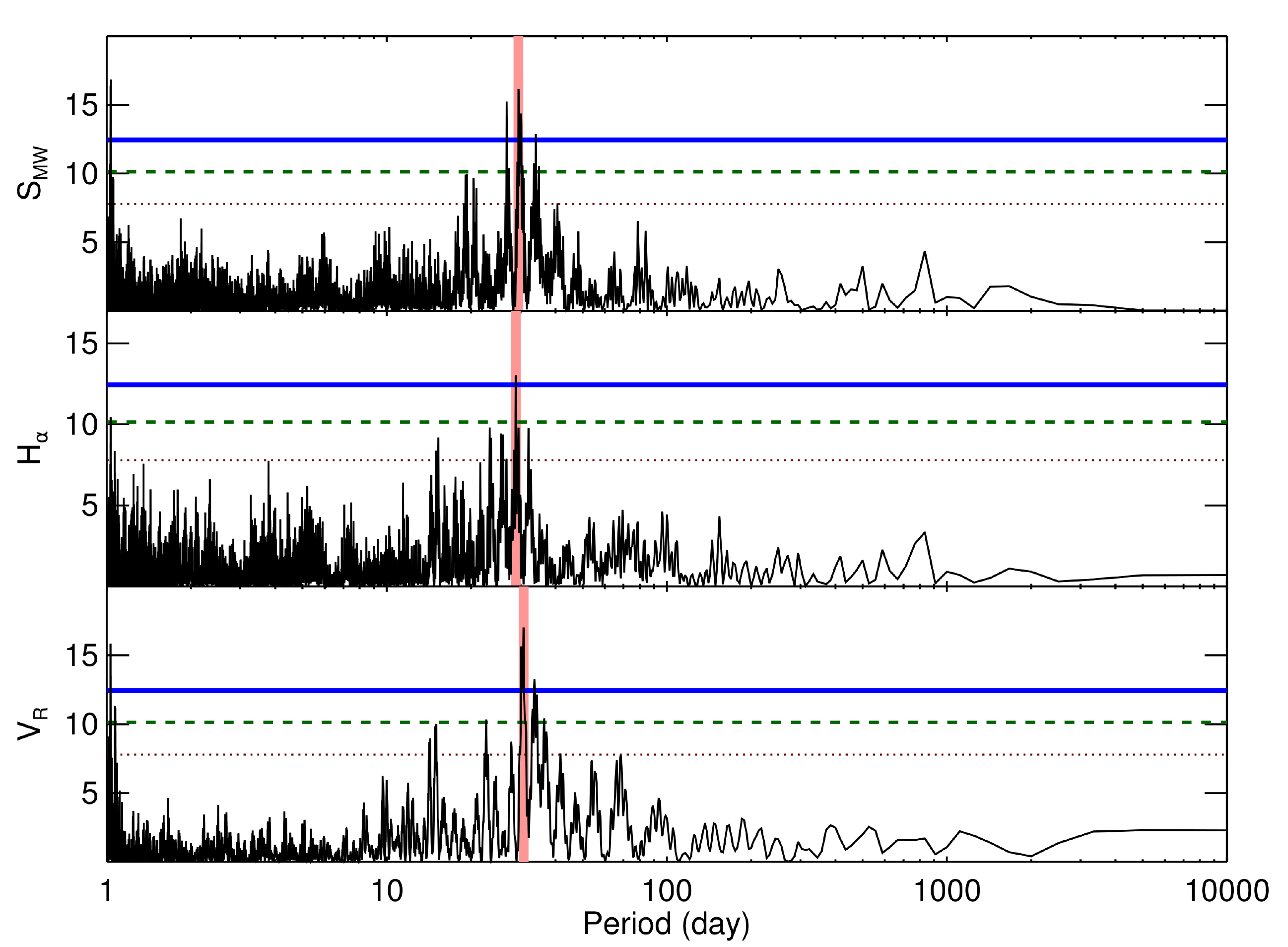}
\caption{Periodograms for the S$_{MW}$ index,  H$_{\alpha}$ index and radial velocity time-series for the M-dwarf GJ 514. Light red fill identifies the rotational modulation. Horizontal lines show the FAP levels. Red dotted line shows the 10\% FAP level, green dashed line the 1\% level and blue solid line the 0.1\% level.}
\label{GJ514_data}
\end{figure}

\begin{figure*}
\begin{minipage}{0.33\textwidth}
        \centering
        \includegraphics[width=6.cm]{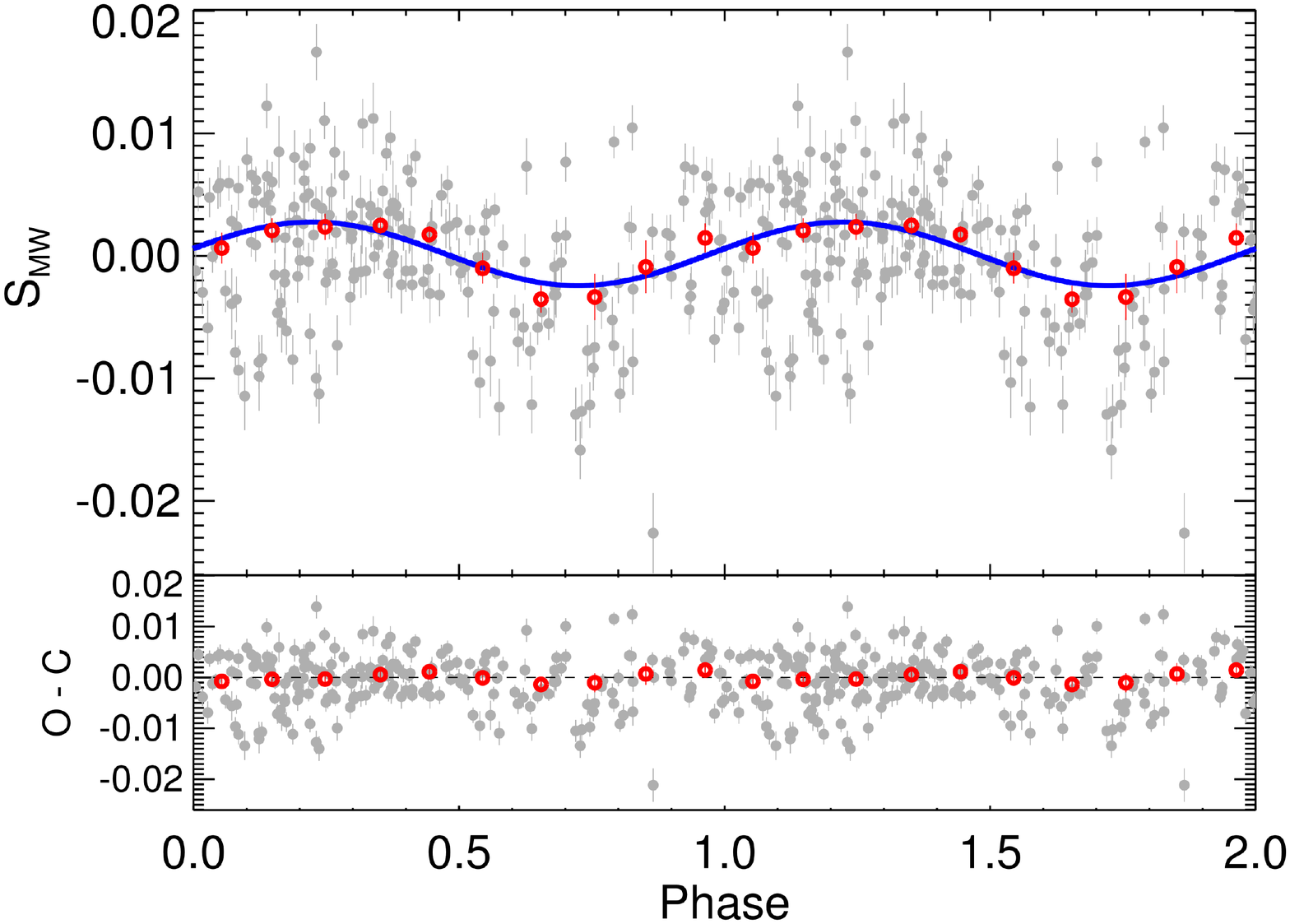}
\end{minipage}%
\begin{minipage}{0.33\textwidth}
        \centering
        \includegraphics[width=6.cm]{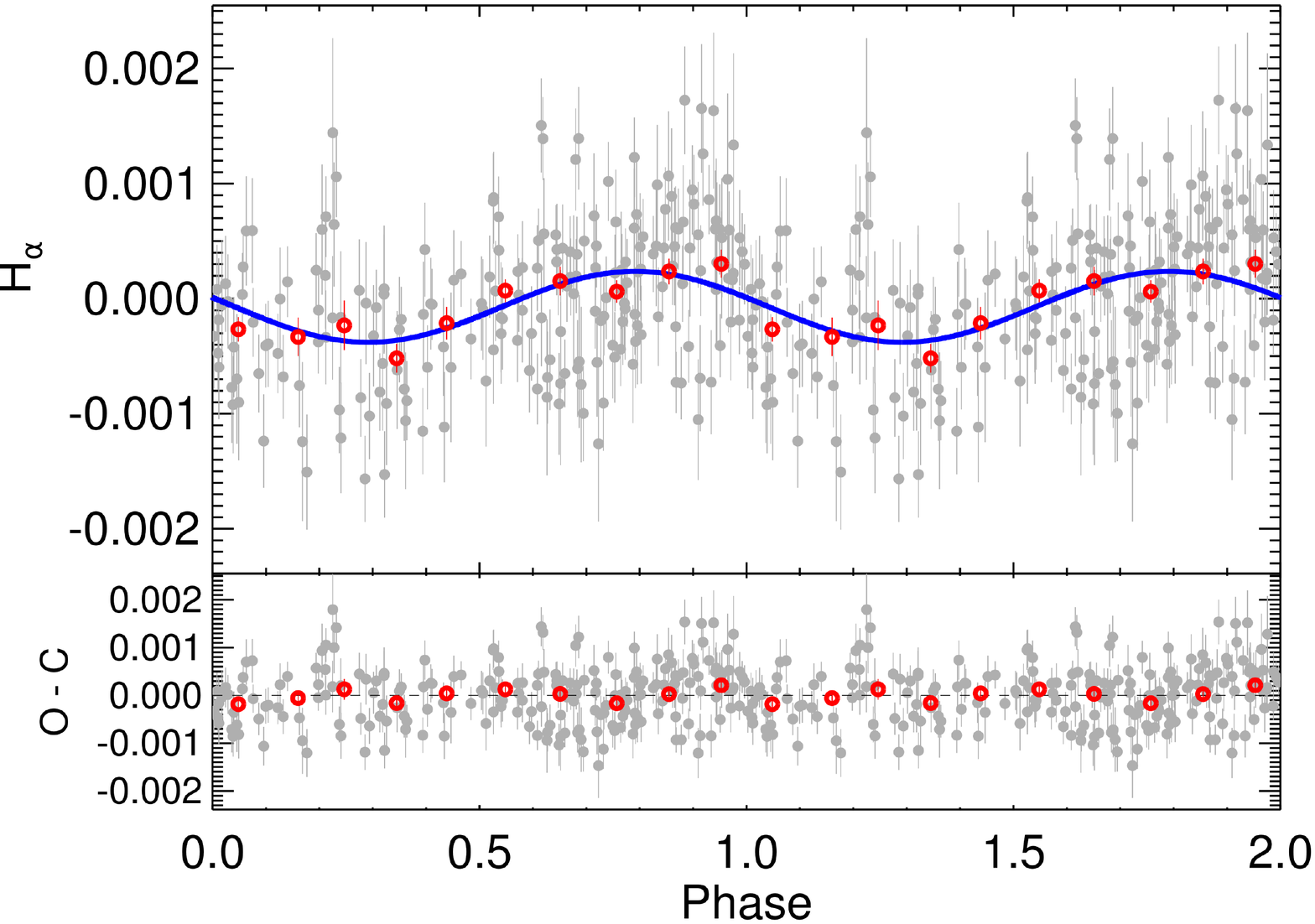}
\end{minipage}%
\begin{minipage}{0.33\textwidth}
        \centering
		\includegraphics[width=6.cm]{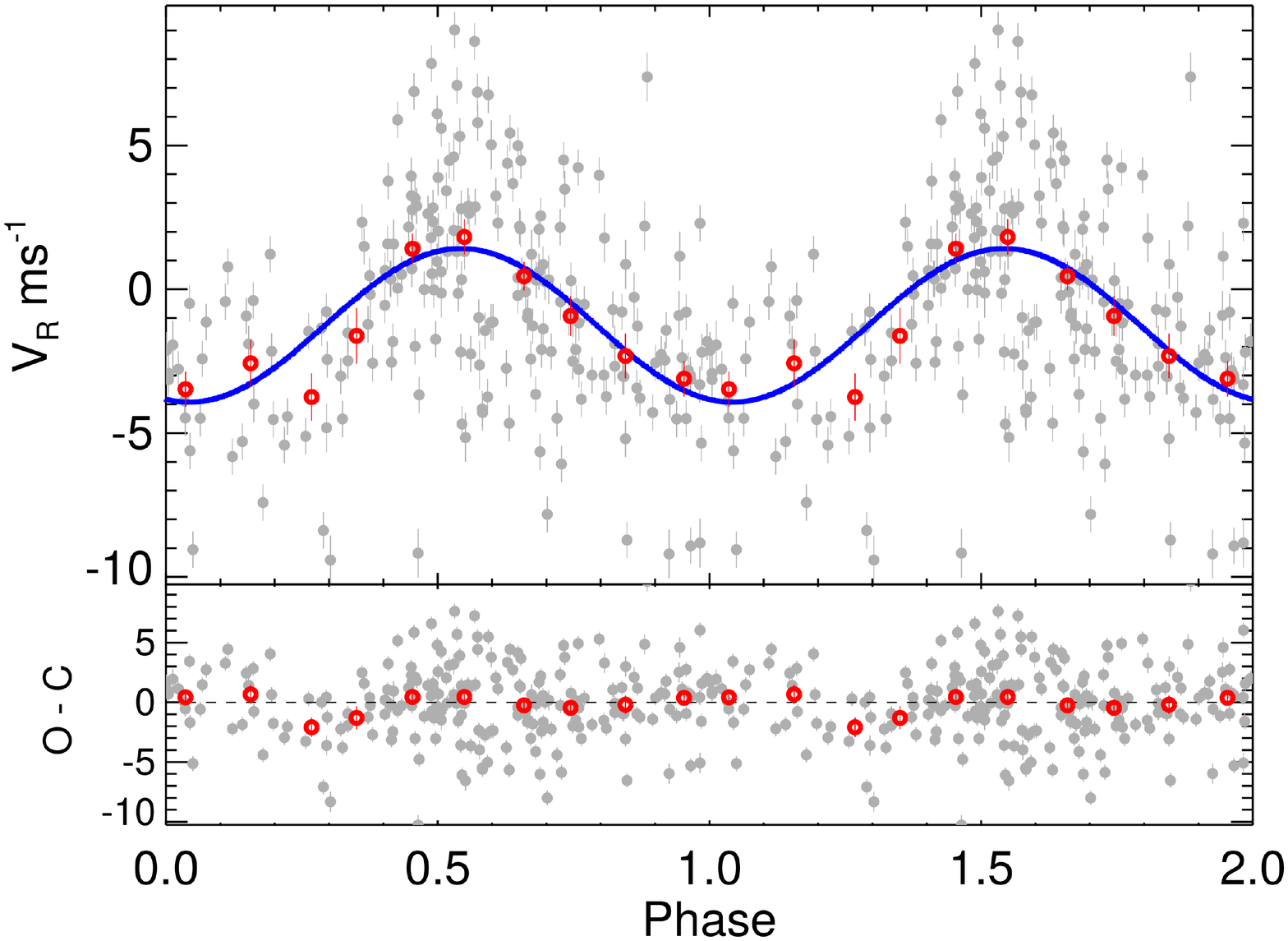}
\end{minipage}%

\begin{minipage}{0.33\textwidth}
        \centering
        \includegraphics[width=6.5cm]{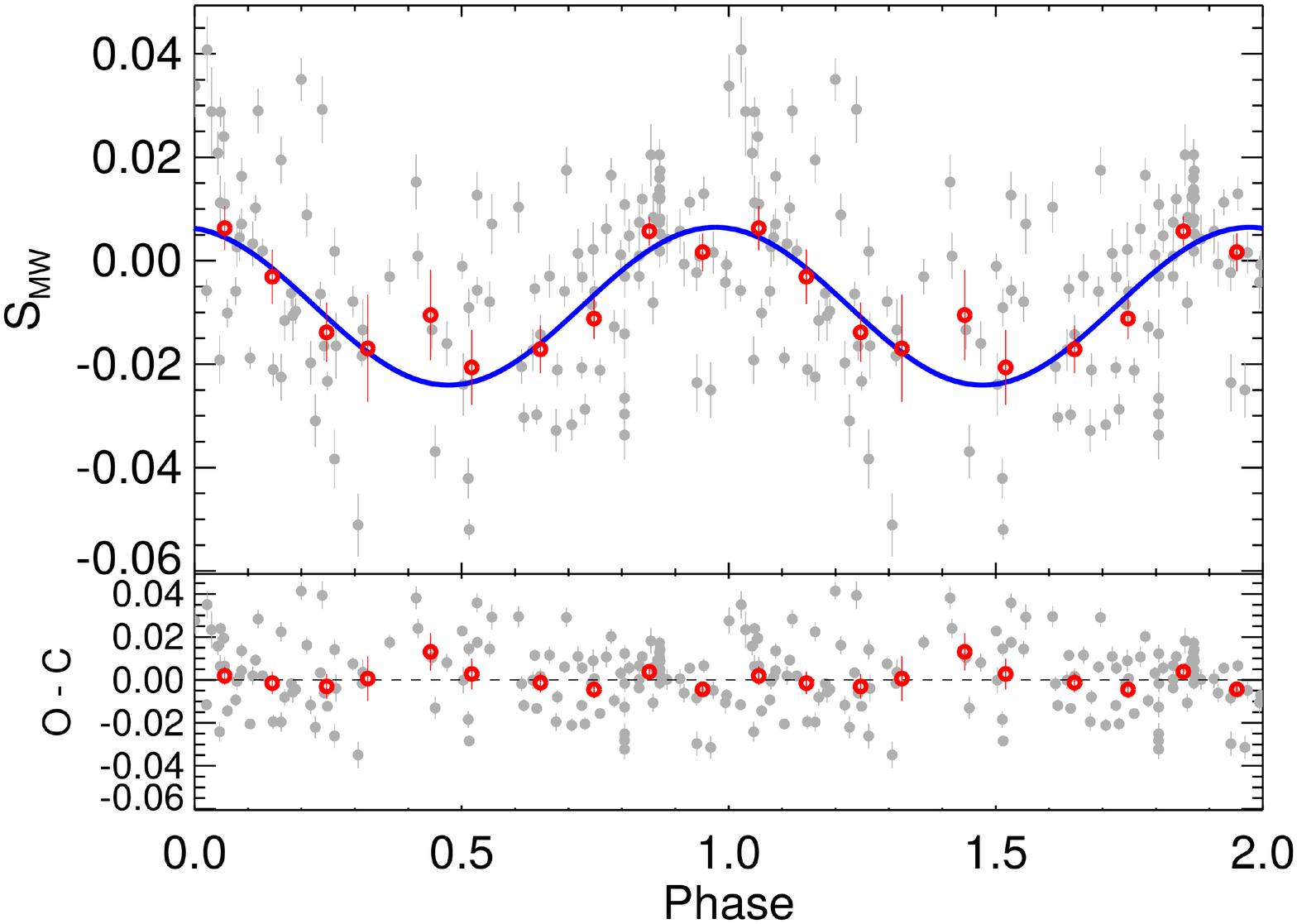}
\end{minipage}%
\begin{minipage}{0.33\textwidth}
        \centering
        \includegraphics[width=6.cm]{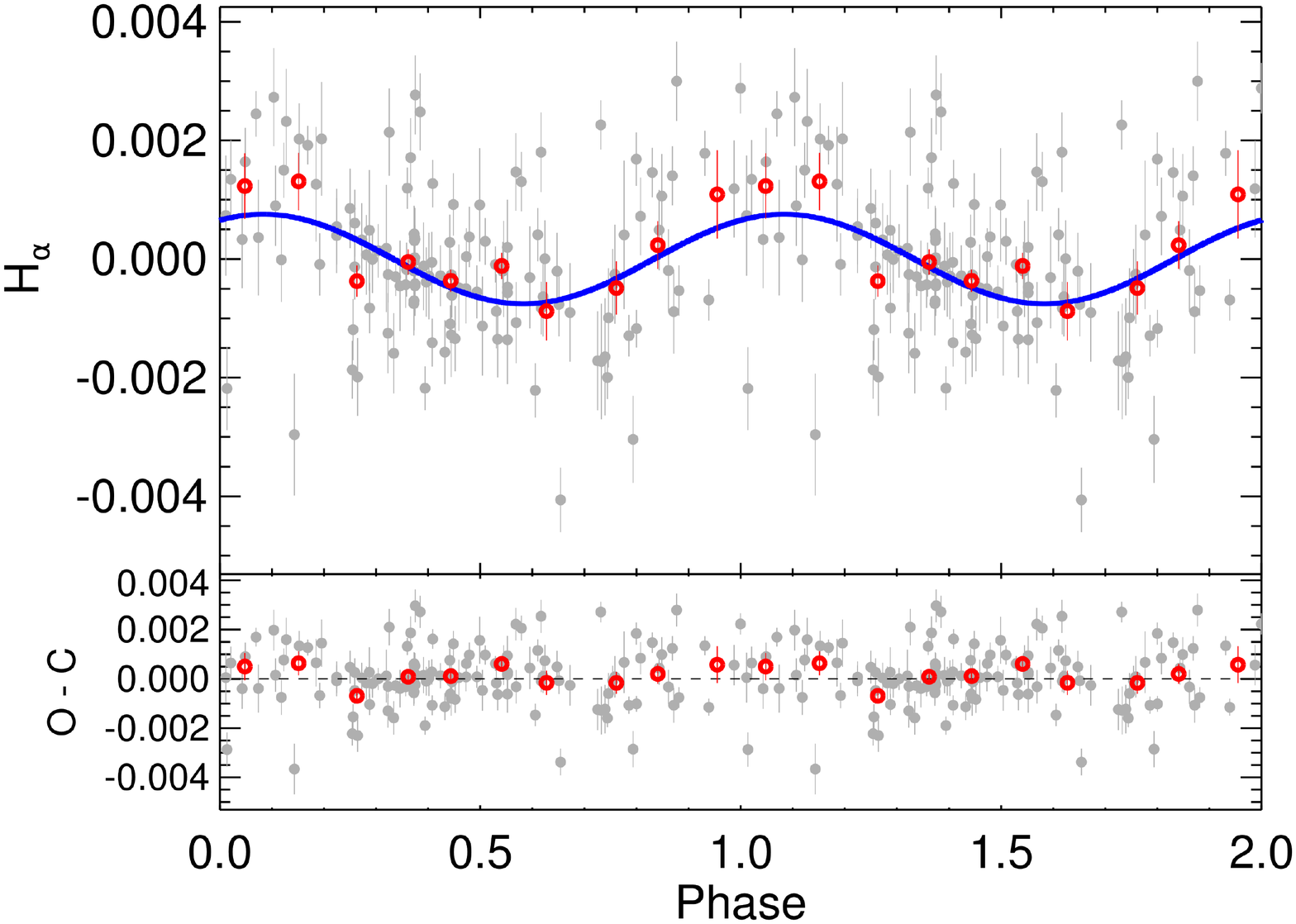}
\end{minipage}%
\begin{minipage}{0.33\textwidth}
        \centering
		\includegraphics[width=6.5cm]{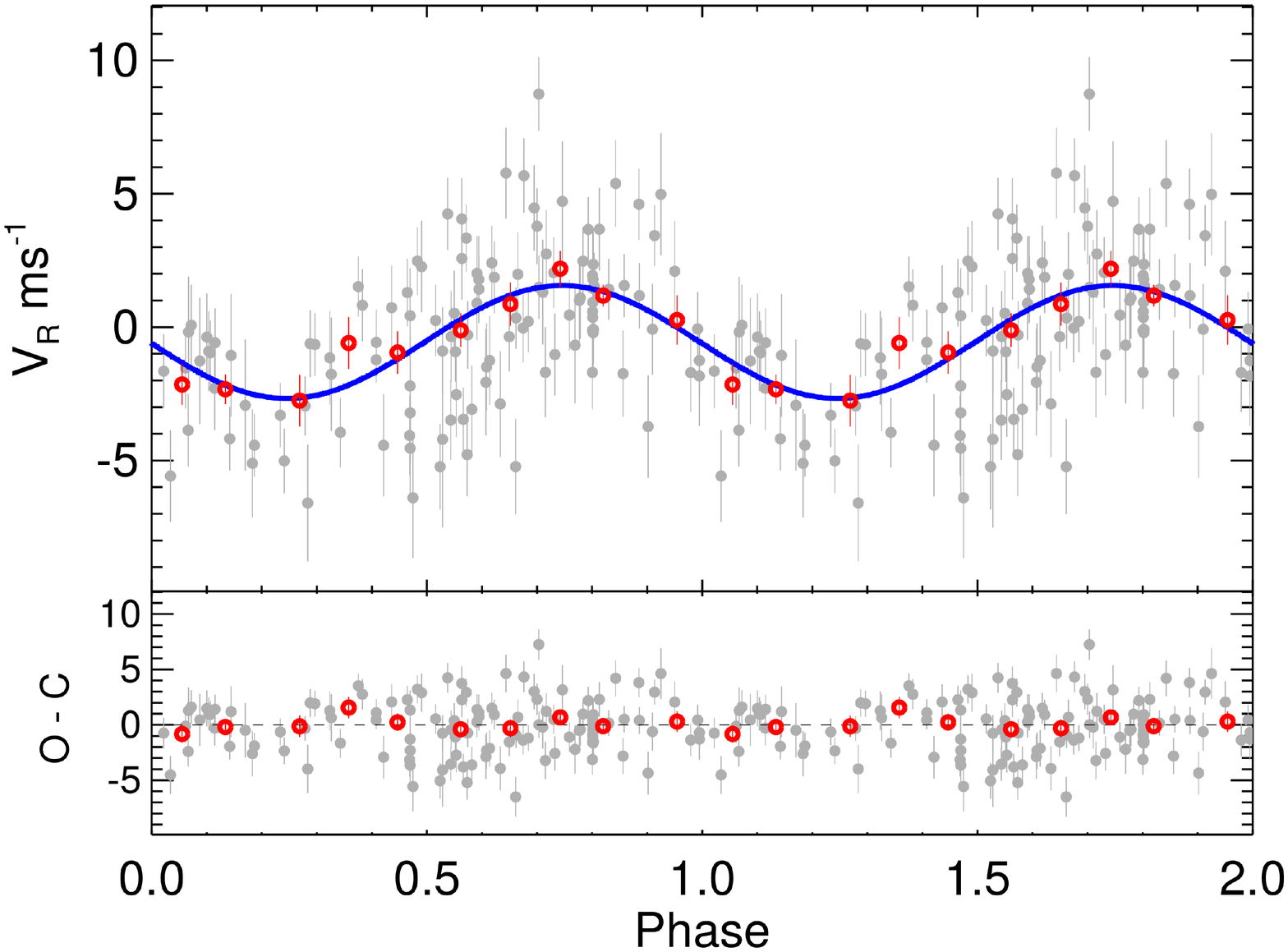}
\end{minipage}%

\begin{minipage}{0.33\textwidth}
        \centering
        \includegraphics[width=6.5cm]{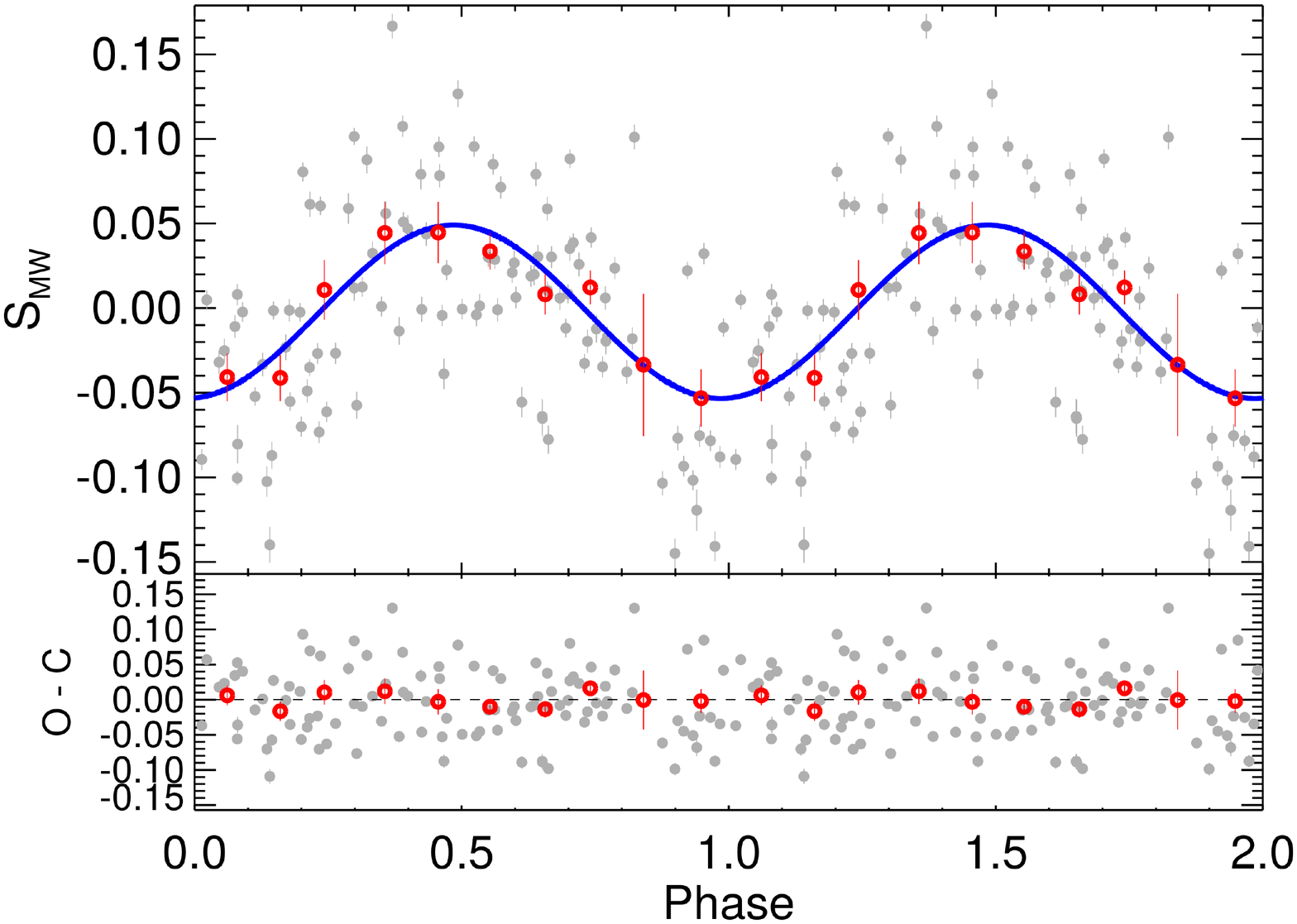}
\end{minipage}%
\begin{minipage}{0.33\textwidth}
        \centering
        \includegraphics[width=6.cm]{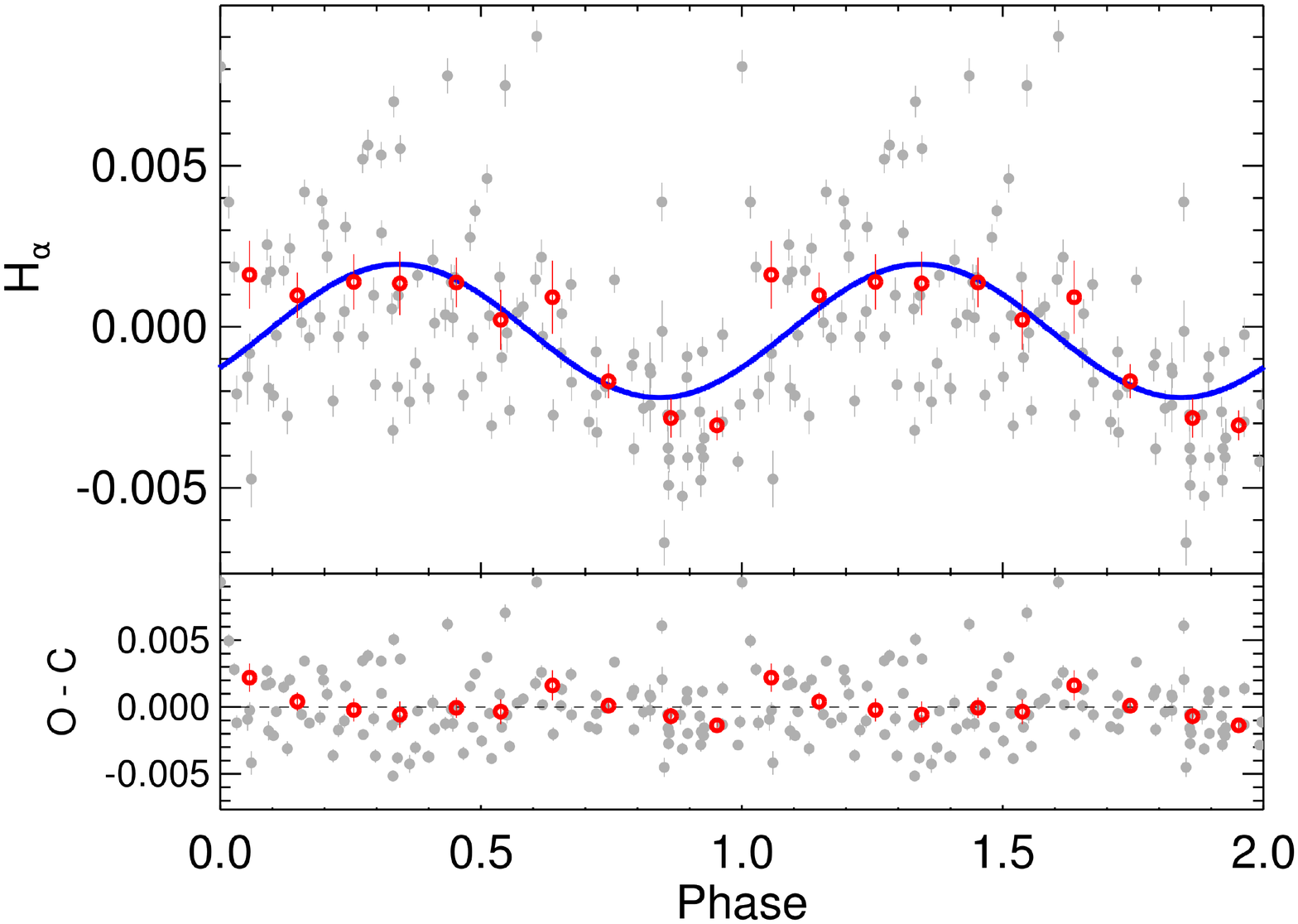}
\end{minipage}%
\begin{minipage}{0.33\textwidth}
        \centering
		\includegraphics[width=6.5cm]{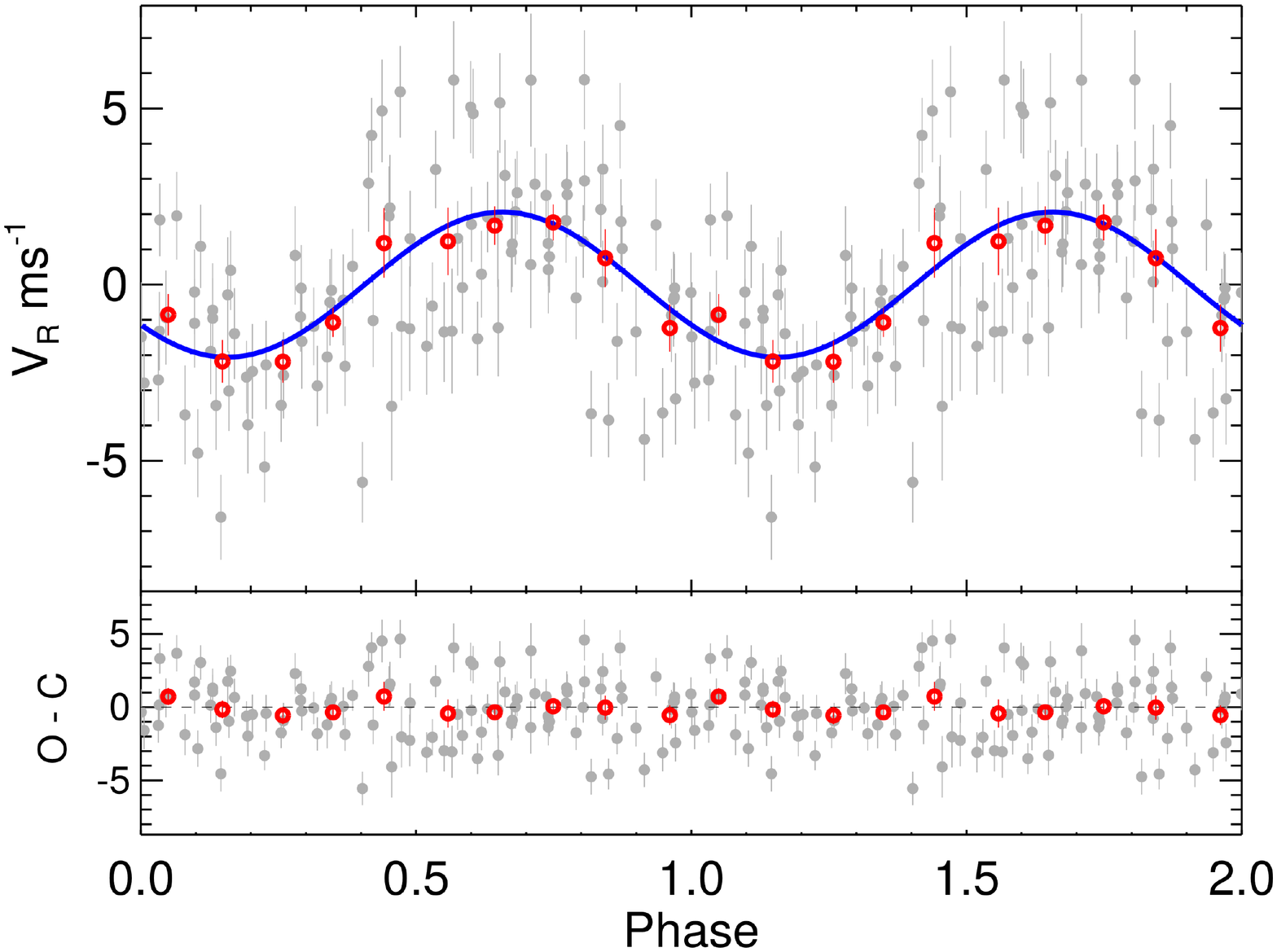}
\end{minipage}%
\caption{Phase folded curve using the rotational modulation for S$_{MW}$ index (left), H$_{\alpha}$ index (center) and radial velocity (right) for the G-type star HD 41248 (top panels), the K-type star HD 125595 (middle panels) and the M-dwarf GJ 514 (bottom panels). Grey dots are the raw measurements after subtracting the mean value. Red dots are the same points binned in phase with a bin size of 0.1. The error bar of a given bin is estimated using the weighted standard deviation of binned measurements divided by the square root of the number of measurements included in this bin. This estimation of the bin error bars assumes white noise, which is justified by the binning in phase, which regroups points that are uncorrelated in time. }
\label{GJ514_fits}
\end{figure*}

Figure~\ref{HD41248_data} shows the periodograms for the $_{MW}$ index,  H$_{\alpha}$ index and radial velocity time-series for the G-type star HD 41248. The three periodograms have a common periodicity which we interpret as the rotation signature of the star in the spectroscopic data. We detect a mean rotation of 23.8 days and a rotation induced radial velocity signal of 25.6 days with a semi amplitude of 2.43 m s$^{-1}$. For the K-type star HD 125595 Figure~\ref{HD125595_data} shows a very similar picture. There is a common periodicity across the different indicators which marks the rotational modulation. We measure a mean rotation period of 38.7 days and a rotation induced radial velocity signal of 36.7 days with a semi amplitude of 2.11 m s$^{-1}$. Figure~\ref{GJ514_data} shows the same scenario for the case of the M-dwarf star GJ 514. There is a common periodicity in the periodograms for the S$_{MW}$ index,  H$_{\alpha}$ index and radial velocity time-series. Once again we interpret this periodicity as the rotation signal of the star. We measure a mean rotation period of 30.0 days and a rotation induced RV signal of 30.8 days with a semi amplitude of 2.06 m s$^{-1}$. Figure~\ref{GJ514_fits} shows the phase folded fits of the three time-series of the three stars using the detected periodicities.

\begin{figure}
\includegraphics[width=9cm]{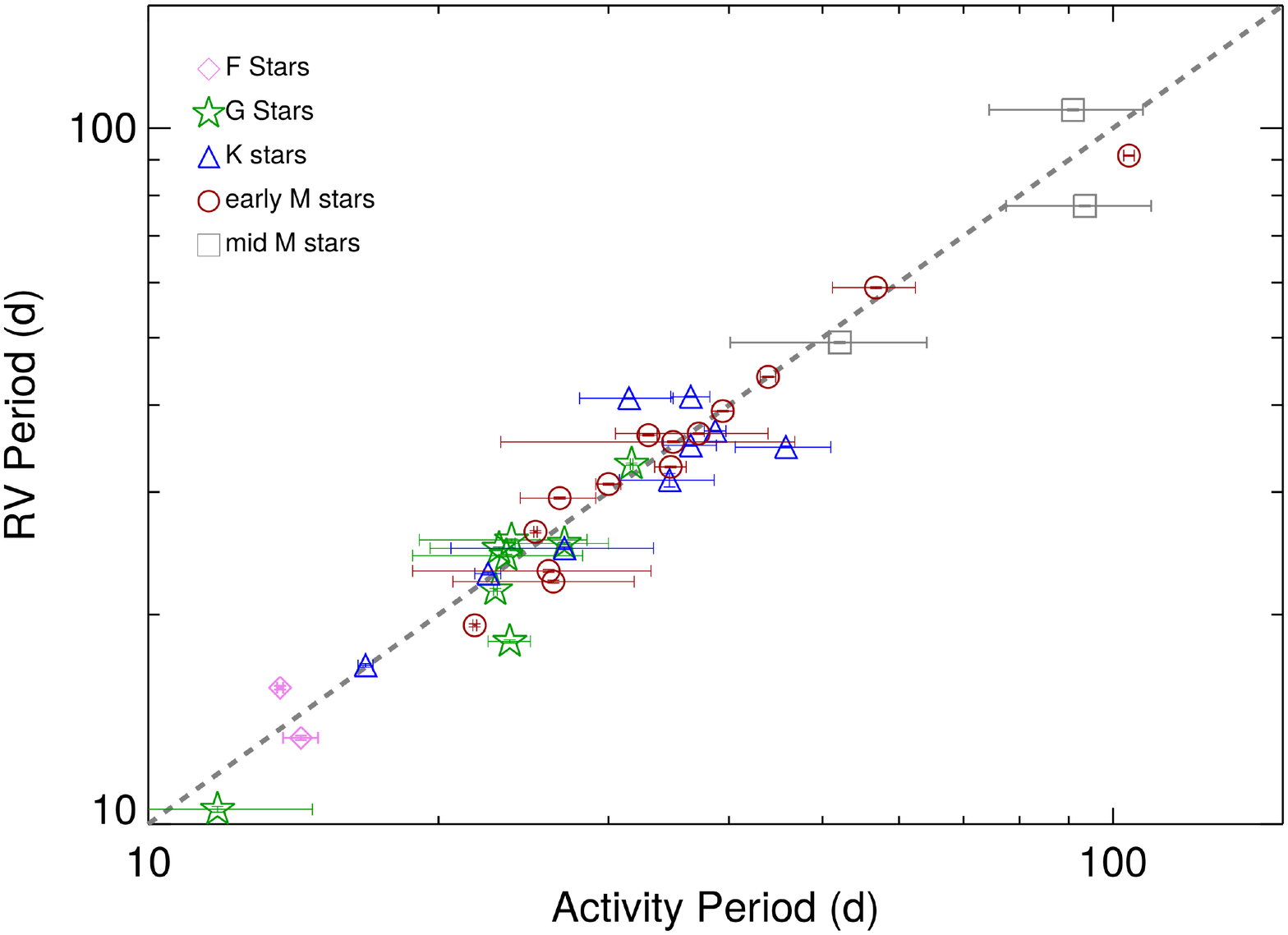}
\caption{Period signal activity induced RV signal versus the rotation period determined using the activity proxies. The grey dashed line shows the 1-1 relationship.}
\label{Rot_com}
\end{figure}

By following this procedure for each star we are able to find rotation induced RV signals in 37 stars of our sample. Table~\ref{table_results} and figure ~\ref{Rot_com} show the results.

\begin {table*}
\begin{center}
\caption { Rotation signals and radial velocity semi-amplitudes for those signals\label{tab:RotationAmp}}
    \begin{tabular}{ l  c  c  c  c c c c c c c } \hline
Name   &   $\log_{10}(R'_\textrm{HK})$ & Rot. Period$^{a}$  & FAP & V$_{r}$ Period$^{b}$ & FAP & V$_{r}$ Amplitude & Confirmed Planets$^{c}$ \\
&   &  (d) &\% & (d) & \% & $(ms^{-1})$ \\ \hline

HD 25171 &   --4.90  $\pm$  0.06  &  14.4  $\pm$  0.6 &   0.8 &   13.6  $\pm$  0.1  &  23.6 &   1.92 $\pm$   0.41 & 1/1  \\
HD 1581 &   --4.84  $\pm$  0.01  &  13.7  $\pm$  0.1 &   $\textless$ 0.01 &   15.7  $\pm$  0.1  &  $\textless$ 0.1 &   0.99 $\pm$   0.23 &   \\
\\

HD 134060 &   --4.91  $\pm$  0.01  &  23.7  $\pm$  1.2 &   $\textless$ 0.1 &   18.3  $\pm$  0.1  &  $\textless$ 0.1 &   0.91 $\pm$   0.07 &  2/2 \\
HD 1388 &   --4.89  $\pm$  0.01  &  22.9  $\pm$  0.1 &   0.2 &   21.7  $\pm$  0.1  &  $\textless$ 0.1 &   1.08 $\pm$   0.12 &   \\
HD 30495 &   --4.50  $\pm$  0.03  &  11.8  $\pm$  3.0 &   $\textless$ 0.1 &   10.5  $\pm$  0.1  &  $\textless$ 0.1 &   7.05 $\pm$   0.16 &   \\
HD 2071 &   --4.83  $\pm$  0.03  &  23.5  $\pm$  4.7 &   $\textless$ 0.1 &   24.3  $\pm$  0.1  &  1.2 &   1.92 $\pm$   0.16 &   \\
HD 41248 &   --4.80  $\pm$  0.03  &  23.8  $\pm$  4.7 &   $\textless$ 0.1 &   25.6  $\pm$  0.1  &  $\textless$ 0.1 &   2.43 $\pm$   0.13 &   \\
HD 1461 &   --4.95  $\pm$  0.02  &  31.7  $\pm$  0.1 &   $\textless$ 0.1 &   32.9  $\pm$  0.1 &  $\textless$ 0.1 &   0.67 $\pm$   0.06 &  2/2 \\
HD 59468 &  --4.95 $\pm$ 0.02  &  23.1  $\pm$  3.5 & $\textless$ 0.1 & 24.9 $\pm$ 0.1 & $\textless$ 0.1 & 0.51 $\pm$ 0.09\\
HD 63765 &   --4.77  $\pm$  0.05  &  27.0  $\pm$  3.0 &   $\textless$ 0.1 &   25.3  $\pm$  0.1  &  1.7 &   3.34 $\pm$   0.18 & 1/1  \\

\\
Corot-7 &   --4.62  $\pm$  0.07  &  22.5  $\pm$  0.7 &   $\textless$ 0.1 &   22.9  $\pm$  0.1  &  $\textless$ 0.1 &   7.53 $\pm$   0.25 &  2/2 \\
Alpha Cen B	& --4.93 $\pm$ 0.04 & 34.7 $\pm$   3.9  & $\textless$ 0.1 &  31.2 $\pm$ 0.7 &$\textless$ 0.1 &  1.81 $\pm$ 0.10 \\
HD 224789 &   --4.43  $\pm$  0.02  &  16.8  $\pm$  0.3 &   $\textless$ 0.1 &   16.9  $\pm$  0.1  &  1.5 &   16.40 $\pm$   0.24 &   \\
HD 40307 &   --5.05  $\pm$  0.05  &  36.5  $\pm$  2.3 &   $\textless$ 0.1 &   35.0  $\pm$  0.1  &  $\textless$ 0.1 &   0.50 $\pm$   0.06 &  5/6 \\
HD 176986 &   --4.83  $\pm$  0.03  &  31.5  $\pm$  3.5 &   $\textless$ 0.1 &   40.9  $\pm$  0.1  &  $\textless$ 0.1 &   1.34 $\pm$   0.11 &   \\
HD 215152 &  --4.94  $\pm$  0.05  &  36.5  $\pm$  1.7 &   $\textless$ 0.1 &   41.1  $\pm$  0.1  &  0.9 &   0.51 $\pm$   0.07 & 2/2  \\
HD 125595 &   --4.77  $\pm$  0.03  &  38.7  $\pm$  1.0 &   $\textless$ 0.1 &   36.7  $\pm$  0.1 &  $\textless$ 0.1 &   2.11 $\pm$   0.19 & 1/1  \\
HD 209100  & --4.78 $\pm$ 0.03 & 27.0 $\pm$ 6.4  &$\textless$ 0.1 &  24.9 $\pm$ 0.1 & $\textless$ 0.1 &  1.41 $\pm$ 0.11\\
HD 85512 &   --4.95  $\pm$  0.04  &  45.8  $\pm$  5.2 &   $\textless$ 0.1 &   34.8  $\pm$  0.1  &  $\textless$ 0.1  &   0.30 $\pm$   0.05 &  1/1 \\

\\
GJ 229 &  --4.91 $\pm$ 0.04 & 26.7 $\pm$ 2.4 &  $\textless$ 0.1 & 29.4 $\pm$ 0.1 & $\textless$ 0.1 & 1.29 $\pm$ 0.14 & 1/1   \\
GJ 846 &  --4.81 $\pm$ 0.03 & 26.3 $\pm$ 5.6 & $\textless$ 0.1 & 22.3 $\pm$ 0.1 & $\textless$ 0.1 & 3.42 $\pm$ 0.33\\
GJ 676 A &   --4.96  $\pm$  0.03  &  35.0  $\pm$  11.8 &   $\textless$ 0.1 &   35.4  $\pm$  0.1  &  0.3 &   2.37 $\pm$   0.28 &  3/4 \\
GJ 514 &   --5.12  $\pm$  0.05  &  30.0  $\pm$  0.9 &   $\textless$ 0.1 &   30.8  $\pm$  0.1  &  $\textless$  0.1 &   2.06 $\pm$   0.16 &   \\
GJ 536 &   --5.12  $\pm$  0.04  &  43.9  $\pm$  0.8 &   $\textless$ 0.1 &   43.9  $\pm$  0.1  &  $\textless$ 0.1  &   2.26 $\pm$   0.92 & 1/1  \\
GJ 205 &  --4.75  $\pm$  0.03  &  34.8  $\pm$  1.3 &   $\textless$ 0.1 &   32.6  $\pm$  0.1  &  $\textless$ 0.1 &   5.97 $\pm$   0.20 &   \\
GJ 667 C &   --5.61  $\pm$  0.07  &  103.9  $\pm$  1.3 &   $\textless$ 0.1 &   91.3  $\pm$  0.2  &  $\textless$ 0.1 &   1.42 $\pm$   0.17 & 5/6  \\
GJ 1 &   --5.53  $\pm$  0.06  &  56.8  $\pm$  5.6 &   $\textless$ 0.1 &   59.0  $\pm$  0.2  &  11.1 &   1.67 $\pm$   0.28 &   \\
GJ 382 &   --4.82  $\pm$  0.04  &  21.8  $\pm$  0.1 &   $\textless$ 0.1 &   19.3  $\pm$  0.1  &  12.3 &   5.69 $\pm$   0.38 &   \\
GJ 832 &  --5.23  $\pm$  0.06  &  39.2  $\pm$  9.4 &   $\textless$ 0.1 &   39.1  $\pm$  0.1  &  0.6 &   1.59 $\pm$   0.22 &  1/2 \\
GJ 880 &   --4.92  $\pm$  0.04  &  37.2  $\pm$  6.7 &   $\textless$ 0.1 &   36.4  $\pm$  0.1  &  $\textless$ 0.1 &   2.93 $\pm$   0.19 &   \\
GJ 176 &   --5.00  $\pm$  0.04  &  39.4  $\pm$  1.0 &   $\textless$ 0.1 &   39.2  $\pm$  0.1  &  $\textless$ 0.1 &   4.01 $\pm$   0.27 &  1/1 \\
GJ 358 &   --4.64  $\pm$  0.03  &  25.2  $\pm$  0.1 &   $\textless$ 0.1 &   26.3  $\pm$  0.1  &  0.4 &   8.64 $\pm$   0.42 &   \\
GJ 479 &  --4.85  $\pm$  0.04  &  26.0  $\pm$  7.2 &   0.6 &   23.1  $\pm$  0.1  &  $\textless$ 0.1 &   4.24 $\pm$   0.30 &   \\
GJ 674 &   --5.04  $\pm$  0.07  &  33.0  $\pm$  0.7 &   $\textless$ 0.1 &   36.2  $\pm$  0.1  &  $\textless$ 0.1 &   2.87 $\pm$   0.15 &  1/1 \\
GJ 273 &   --5.44  $\pm$  0.05  &  93.5  $\pm$  16.0 &   $\textless$ 0.1 &   77.3  $\pm$  0.1  &  $\textless$ 0.1 &   1.30 $\pm$   0.13 &   \\
GJ 526 &  --5.06  $\pm$  0.04  &  52.1  $\pm$  12.0 &   0.9 &   49.2  $\pm$  0.1  &  6.4 &   3.81 $\pm$   0.38 &   \\
GJ 876 &   --5.30  $\pm$  0.05  &  90.9  $\pm$  16.5 &   $\textless$ 0.1 &   106.2  $\pm$  0.2  &  $\textless$ 0.1 &   1.99 $\pm$   0.12 & 4/4  \\
 \hline
\label{table_results}
\end{tabular} 
\end{center}
\begin{flushleft}
$^{a}$ Weighted mean measurement of all the individual measurements in the different activity proxies. Errors in the determination are the standard deviation of those measurements.  \\
$^{b}$ Errors in the V$_{r}$ period are the 1-$\sigma$ errors given by the least squares minimization process.\\
$^{c}$ Number of planet candidates confirmed out of the number of published planet candidates. For the cases of HD 40307 e, GJ 676 A e, GJ 667 C d and GJ 832 b we detect the claimed signal, but interpret it as a rotation induced signal. 
\end{flushleft}

\end{table*}

There could be multiple reasons why we cannot find  the rotation induced RV signals in the remaining 17 stars of the sample. In some situations -- low activity stars, specially in the case of faint stars -- the RV induced signal might be below the noise level. In some others, the geometric pattern of activity features might cause signals which are not at the rotation period of the star, but at one of its harmonics. The inclination of the stellar axis related to our line of sight might also play a role in the relative proportion between harmonics and the shape of the RV induced signal by altering the apparent geometric pattern. Even minimizing the amplitude to an undetectable degree in the case of stars aligned pole-on. Short-lived activity regions could also undermine the detectability of the signals. In order to measure a coherent signal it is important for the active regions to last at least a few rotational periods. It has been seen that, in the case of some M-type stars, active regions are stable over long time spans \citep{Robertson2015}, but it does not have to be true for all stars.

\subsection{Radial velocity signals induced at the harmonics of the rotation period}

When measuring rotation induced RV signals it is expected to find also modulation with periods around the first harmonics of the rotation period. The relative strength of each of the present signals depends on the surface configuration and the inclination of the rotation axis related to our line of sight. For some particular configurations the signals at the harmonics of the rotation are dominant \citep{Boisse2011}. 

We performed an analysis equivalent to the previous searching for radial velocity signals at the harmonics of the rotation period. A signal at one of these harmonics would be considered an activity induced signal if there is an equivalent signal in any of the available activity indicators or a strong correlation between the radial velocity signal and at least one activity indicator. 

We found activity induced signals associated to harmonics of the rotation period in 9 stars, 8 of them with FAPs smaller than 1\%. Table~\ref{harmonics} shows the periods and semi-amplitudes of the detected signals. Only one of the detections corresponds to a star not included in Table~\ref{table_results}. The star GJ 581 shows an induced signal at half the rotation period but not at the rotation period. Of the 8 stars where we detect a signal both at the rotation period and at one of its harmonics, only one star shows the signal larger for the harmonic. The star HD1461 shows a 0.49 ms$^{-1}$ signal at the rotation period and  0.72 ms$^{-1}$ at the harmonic. In the 7 remaining stars the RV signal at the harmonic is always smaller. 

\begin {table}
\begin{center}
\caption { Radial velocity semi-amplitudes for the rotation induced signals\label{tab:harmonics}}
    \begin{tabular}{ l  c  c  c  c c c c c c c } \hline
Name   & Harm. &  V$_{r}$ Period  &   V$_{r}$ Amp. & FAP \\
&   &  (d)  & $(ms^{-1})$ &\%\\ \hline
HD 1581  &  P/2 & 6.4 $\pm$ 0.1 & 0.34 $\pm$ 0.04 &  $\textless$ 0.1   \\ \\
HD 134060  &  P/2 & 10.4 $\pm$ 0.1 & 0.78 $\pm$ 0.08 &  $\textless$ 0.1  \\
HD 41248  &  P/2 & 13.4 $\pm$ 0.1 & 1.76 $\pm$ 0.14 &  $\textless$ 0.1 \\
HD 1461  &  P/2 & 15.0 $\pm$ 0.1 & 0.72 $\pm$ 0.02 & $\textless$ 0.1  \\ \\
Corot-7  &  P/2 & 11.0 $\pm$ 0.1 & 3.84 $\pm$ 0.23 & $\textless$ 0.1  \\
 &  P/3 & 7.3 $\pm$ 0.1 & 4.28 $\pm$ 0.24 & $\textless$ 0.1  \\ \\
GJ 514 & P/2 & 15.2 $\pm$ 0.1 & 1.55 $\pm$ 0.18 & $\textless$ 0.1 \\
GJ 205 & P/3 &  11.8 $\pm$ 0.1 & 1.92 $\pm$ 0.21 & 0.2 \\
GJ 358  &  P/2 & 13.1 $\pm$ 0.1 & 3.61 $\pm$ 0.38 & 12.7 \\
GJ 581 & P/2 & 66.9 $\pm$ 0.1 & 1.41 $\pm$ 0.13 &  $\textless$ 0.1\\
 \hline
\label{harmonics}
\end{tabular} 
\end{center}
\begin{flushleft}
\end{flushleft}

\end{table}

\begin{figure}
\includegraphics[width=\linewidth]{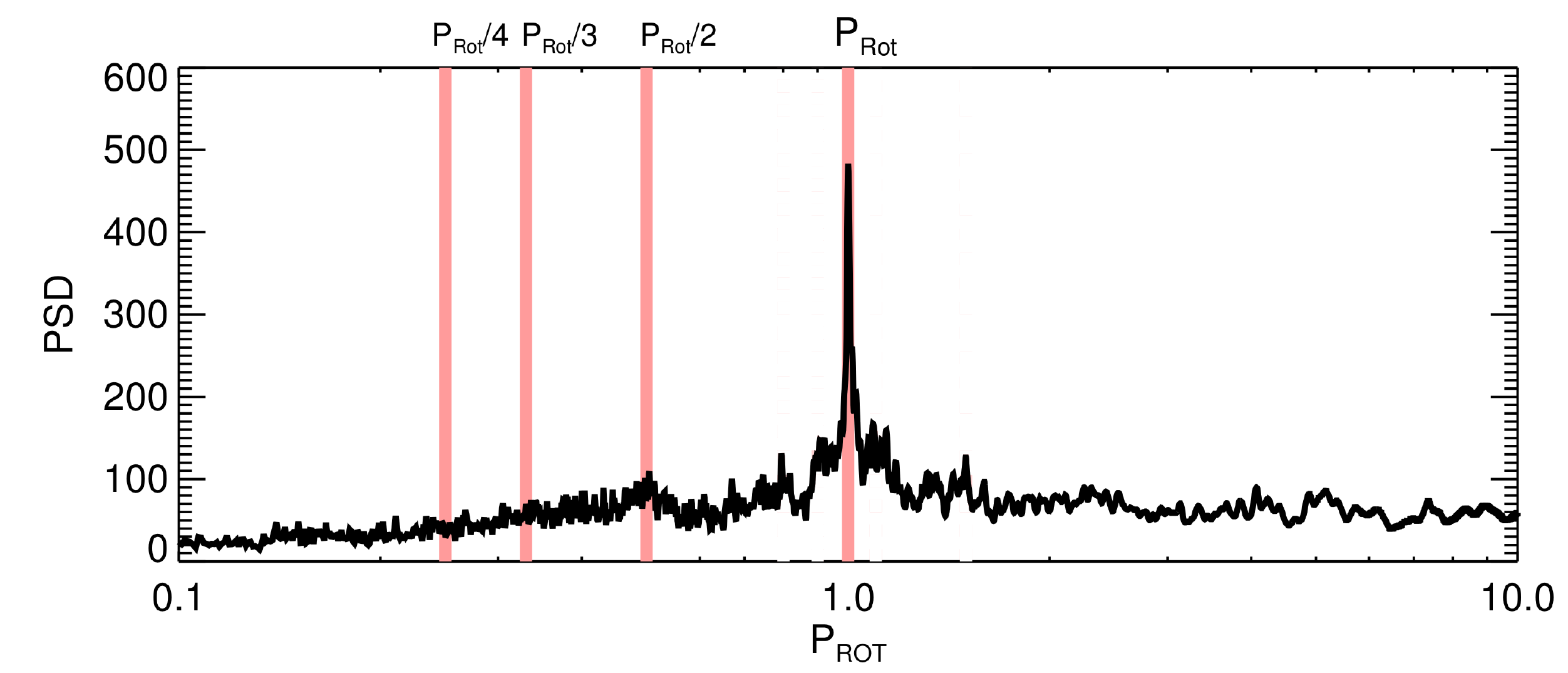}
\caption{Combined periodogram of the radial velocity time-series of all of our stars with detected rotational modulations, after subtracting all planetary signals, with the period axis normalized at the rotation period. Light red fill shows the position of the rotation period and the first harmonics.}
\label{combined_psd}
\end{figure}

Figure~\ref{combined_psd} shows the combined periodogram for the radial velocities of all the stars from Table~\ref{data_sample}, after removing the planetary signals, with the time axis normalized at the rotation period of each individual star. The rotation signal is the dominant feature in this combined periodogram, being the only clear signal. There is a marginal structure at half the rotation period, telling us once more that some of the stars have signals at the first harmonic of the rotation. No other prominent features are present in our combined periodogram. 

\subsection{Implications for planet-hosting candidates}

In the process of our search for rotation-induced radial velocity signals we recover all the previously published radial velocity signals attributed to extrasolar planets. For most of them we agree on the planetary origin of the signal, but there are four cases  where we think a rotation induced signal is more likely to be causing the radial velocity modulation: The cases of HD 40307  e,  GJ 676 A e, GJ 667C d and GJ 832 c. 

HD 40307  was initially claimed to be the host of 3 super-Earths at $4.3, 9.6$ and $20.4 ~d$ \citep{Mayor2009b}. Then the system was expanded to six planets  \citep{Tuomi2013}, planet e being a small super-Earth at $\sim$34.6 day orbital period. This period coincides within the uncertainties with our stellar rotation detection ($36.5 ~d$) and therefore we ascribe to rotation the radial velocity signal, as we previously pointed in \citet{Masca2015}.

GJ 676 A has 4 detected planet candidates. Planet b detected at $\sim1050 ~d$  \citep{Forveille2011} and planets c at $\sim4400 ~d$, d at $\sim3.6 ~d$ and e at $\sim35.4 ~d$  \citep{AngladaEscude2012}. We recover the signals attributed to planets b, d and e and the trend attributed to planet c. As suggested in \citet{Masca2015}, the RV signal of planet e is most likely caused by stellar rotation.

GJ 667 C is a star that hosts 6 planet candidates. Planets b at $\sim7.2 ~d$ and c at $\sim28.1 ~d$ \citep{Bonfils2013}, and planets d at $\sim91.6 ~d$, e at $\sim62.2 ~d$, f at $\sim39.0 ~d$ and g at $\sim256 ~d$~\citep{Gregory2012, AngladaEscude2013}. \citet{Robertson2014b} and \citet{Feroz2014} claimed that planets d and e are artefacts induced by stellar rotation, while  \citet{Delfosse2013} had already identified the $91 ~d$ signal as an alias of an  $\sim106 ~d$ activity signal.  We support the idea of the signal at $\sim91 ~d$ being an artefact caused by the stellar rotation. We agree on the $\sim106 d$ rotation period and we also found a power excess at $\sim91 ~d$ in the $S_{\rm MW}$ periodogram ~\citep{Masca2015}, supporting the idea that the signal might be the mark of rotation measurements at different latitudes.

GJ 832 hosts 2 potential planets. Planet b at $\sim3416 ~d$ \citep{Bailey2009} and planet c at $\sim35.7 ~d$ \citep{Wittenmyer2014}, the latter with an orbital period very close to our rotation period measurement. As in \citet{Masca2015}, we interpret the signal of planet c as the rotation-induced RV signal \citep[see also][]{Bonfils2013}.

\section{Discussion}

\begin{figure}
\centering
\includegraphics[width=\linewidth]{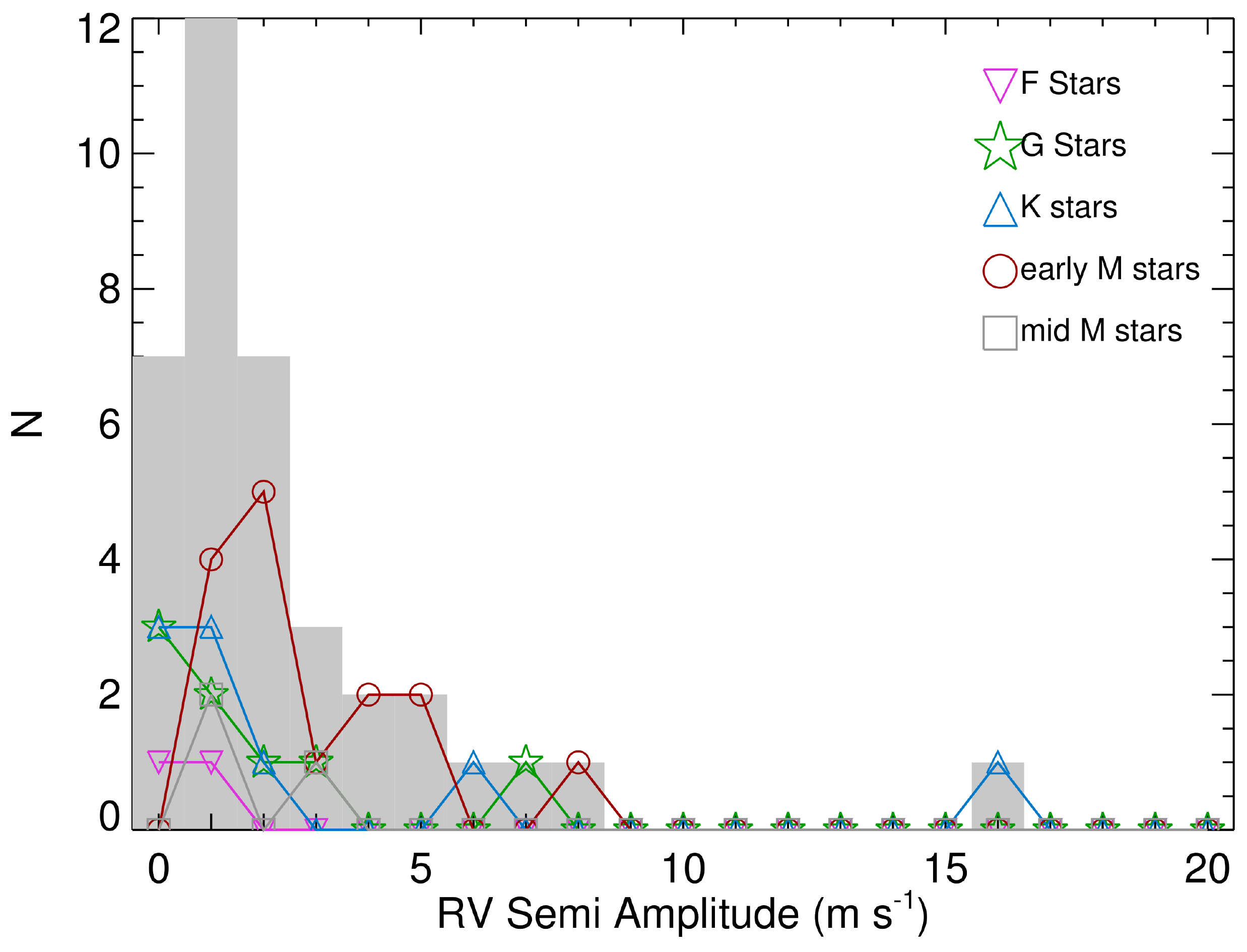}
\caption{Distribution of semi-amplitudes for the rotational modulations of the stars in our sample.}
\label{rv_histo}
\end{figure}

The previous analysis of radial velocity signals induced by stellar rotation provided a collection of 35 stars where we were able to measure the rotation RV signal and/or one of its harmonics, for stars going from late F to mid M-type. For our sample the distribution of radial velocity induced signals peaks at 1-2 m s$^{-1}$. Figure~\ref{rv_histo} shows the distribution of the semi-amplitudes of the periodic RV signal induced by the rotation of the stars for the different spectral types. 

\begin{figure*}
\centering
\includegraphics[width=\textwidth]{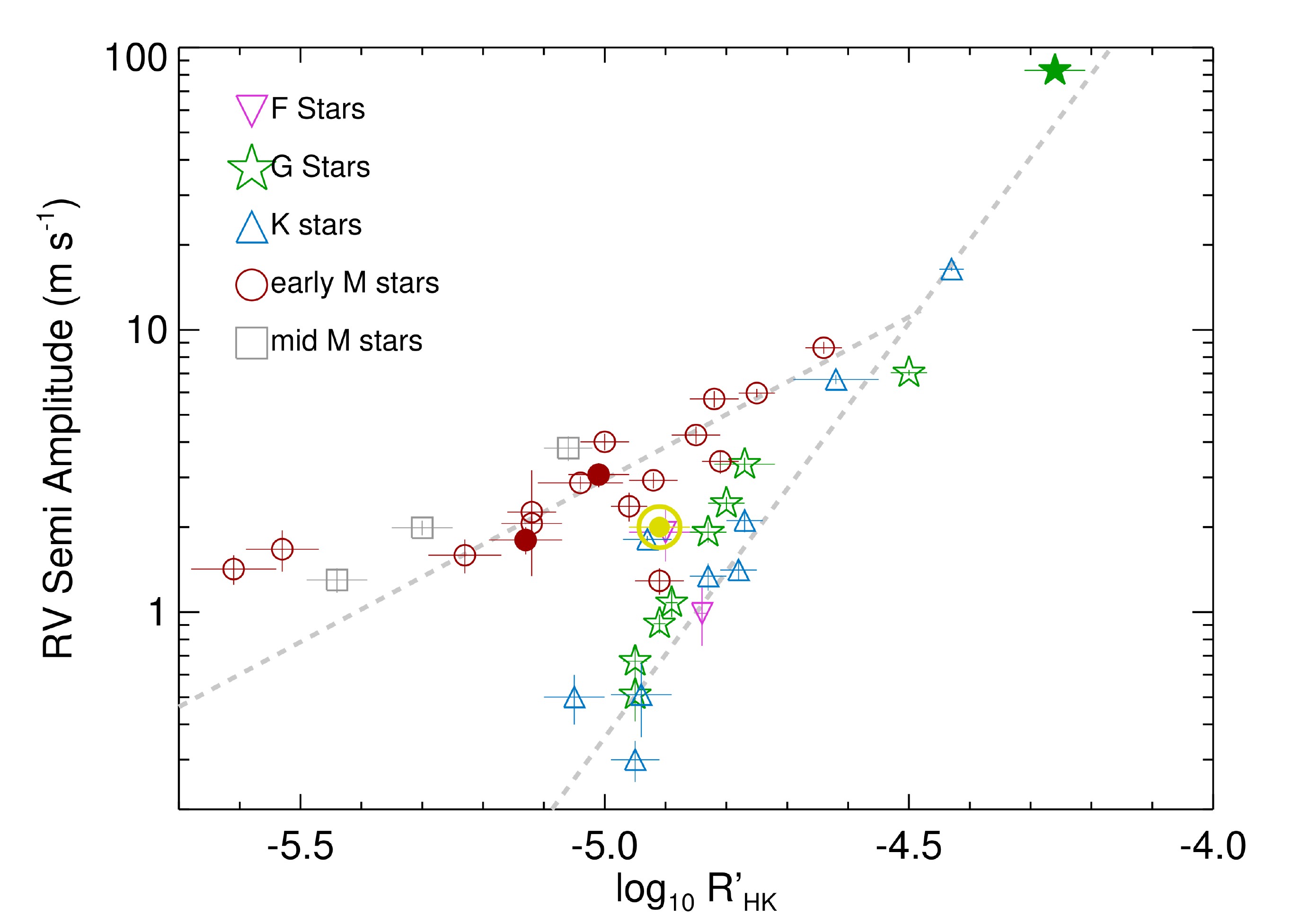}
\caption{Semi-amplitude of the rotation induced signal versus chromospheric activity level $\log_{10}(R'_\textrm{HK})$. Empty symbols correspond to our data, filled symbols the literature data. The grey dashed lines show the best fit to the date for the two groups of stars: FGK-type stars on one group, M-type stars on the other.}
\label{amp_rhk}
\end{figure*}

We found that there is a clear relationship between the activity level of the star and the amplitude of the rotation induced RV signal. Figure~\ref{amp_rhk} shows two linear relationships between the activity level of the star and the logarithm of the amplitude of its induced signal. This relationship is different for M-dwarfs and for FGK stars. We added a few extra points from the literature (Table~\ref{table_results_lit}) in order to populate a bit more the plot and to increase the activity coverage. Both groups of stars are fit independently:
\begin{equation}
   \log_{10}(K)= a \cdot \log_{10}(R'_{\rm HK}) + b,
   \label{eq_rhk}
\end{equation}

Table~\ref{fit_parameters} shows the parameters for the different fits for the two groups of stars. 

We see some discrepancy between our results in Corot-7 and GJ 674 and the results of \citet{Boisse2011} and \citet{Bonfils2007}. Our measurement of the amplitudes for the activity induced RV signals is much smaller. In both cases we are using much longer observation baselines, averaging the amplitude over longer periods of time. In the case of Corot-7 \citet{Boisse2011} performed the analysis during a time when the activity level of the star was higher than the average level of our full dataset, and therefore a larger RV amplitude for the activity induced signal is expected. For the case of GJ 674 the sampling rate of the modern data is much better at the scale pertinent for rotation period measurements than in the original dataset. The amplitude of the rotation induced RV signal might have been overestimated. For HD 41248, $\alpha$~Cen B and GJ 176 our measurements are compatible with those previously published. We included the cases of HD 166435, the Sun, GJ 3998 and GJ 15 A as a comparison. The RV semi amplitude of their rotation signals match quite well the expected value according to their mean activity levels. See Table~\ref{table_results_lit}.

Even in the case of slow rotators the radial velocity signals induced by stellar rotation have semi-amplitudes larger than those of terrestrial planets in the habitable zones. In the case of FGK stars these signals can drop below the 1 ms$^{-1}$ threshold for stars more quiet than the Sun, reaching sometimes the HARPS precision limit. For M-dwarfs even in the case of very slow rotators these signals are larger than 1 ms$^{-1}$. We have not detected any rotation induced RV signal smaler than 1 ms$^{-1}$ in an M-dwarf star. Even in the case of the most quiet stars the radial velocity signal induced by the rotation is still larger than those of terrestrial planets. Modelling and removing these signals becomes necessary.

\begin {table*}
\begin{center}
\caption {Previously published rotation induced radial velocity signals.}
    \begin{tabular}{ l  l  l  c  c c l c c c c } \hline
Name   & SpTp &  $\log_{10}(R'_\textrm{HK})$ & Period  & V$_{r}$ Period & Amplitude  & References & Comments\\
&   & & (d) & (d) & $(ms^{-1})$ \\ \hline
HD 166435 & G1  & --4.26			   & 3.80            & 3.80              & 83.0          & \citet{Queloz2001} & \\
Sun		 & G2  & --4.91             & 24.5             &                   & 2.4             &  \citet{Haywood2016} & \\
HD 41248 & G2  & --4.90              & 20 $\pm$ 3		  & 25.6       & 3.3 $\pm$ 0.26 & \citet{Santos2014}\\ 
Corot-7  & K0  & --4.61              & 23              & 18-23      & & \citet{Queloz2009}\\
		 &		&					& 23				  & 23 & 14.7 $\pm$ 0.8 & \citet{Boisse2011}\\
Alpha Cen B & K1 & --4.94   $\pm$ 0.11 & 38.1 $\pm$ 1.4 &   38.7          & 1.5            & \citet{Dumusque2012} & \\
GJ 3998 & M1 & --5.01  & 30.8 $\pm$ 2.5 & 30.7 & 3.07 $\pm$ 0.3 & \citet{Affer2016}\\
GJ 15 A    & M2 & --5.13 $\pm$ 0.06 &  44.8             & 44.8             & 1.8                & \citet{Howard2014} & 1 \\
GJ 176     & M2.5 & --5.00 $\pm$ 0.04 & 39 & 39 & 4.4 & \citet{Robertson2015} & 1 \\
GJ 674   & M3 & --5.04 $\pm$ 0.07 & 34.8 & 34.8 $\pm$ 0.1 & 5.06 $\pm$ 0.19 & \citet{Bonfils2007} & 2\\

\hline
\label{table_results_lit}
\end{tabular} 
\end{center}
\begin{flushleft}
\textbf{Comments:} 1 - $\log_{10}(R'_\textrm{HK})$ calculated using the Period - $\log_{10}(R'_\textrm{HK})$ relationship from \citet{Masca2015}. 
2 - $\log_{10}(R'_\textrm{HK})$ from this work. \\
\end{flushleft}
\end {table*}

\begin {table}
\begin{center}
\caption {Parameters for equation~\ref{eq_rhk} \label{tab:Parameters}}
    \begin{tabular}{ c  c  c } \hline
Dataset & a & b  \\ \hline
GK-type Stars & 2.93 $\pm$ 0.03 & 14.23 $\pm$ 0.12\\
M-type Stars & 1.15 $\pm$ 0.02 & 6.23 $\pm$ 0.08\\ \hline
\label{fit_parameters}
\end{tabular}  
\end{center}
\end {table}

\begin{figure}
\centering
\includegraphics[width=\linewidth]{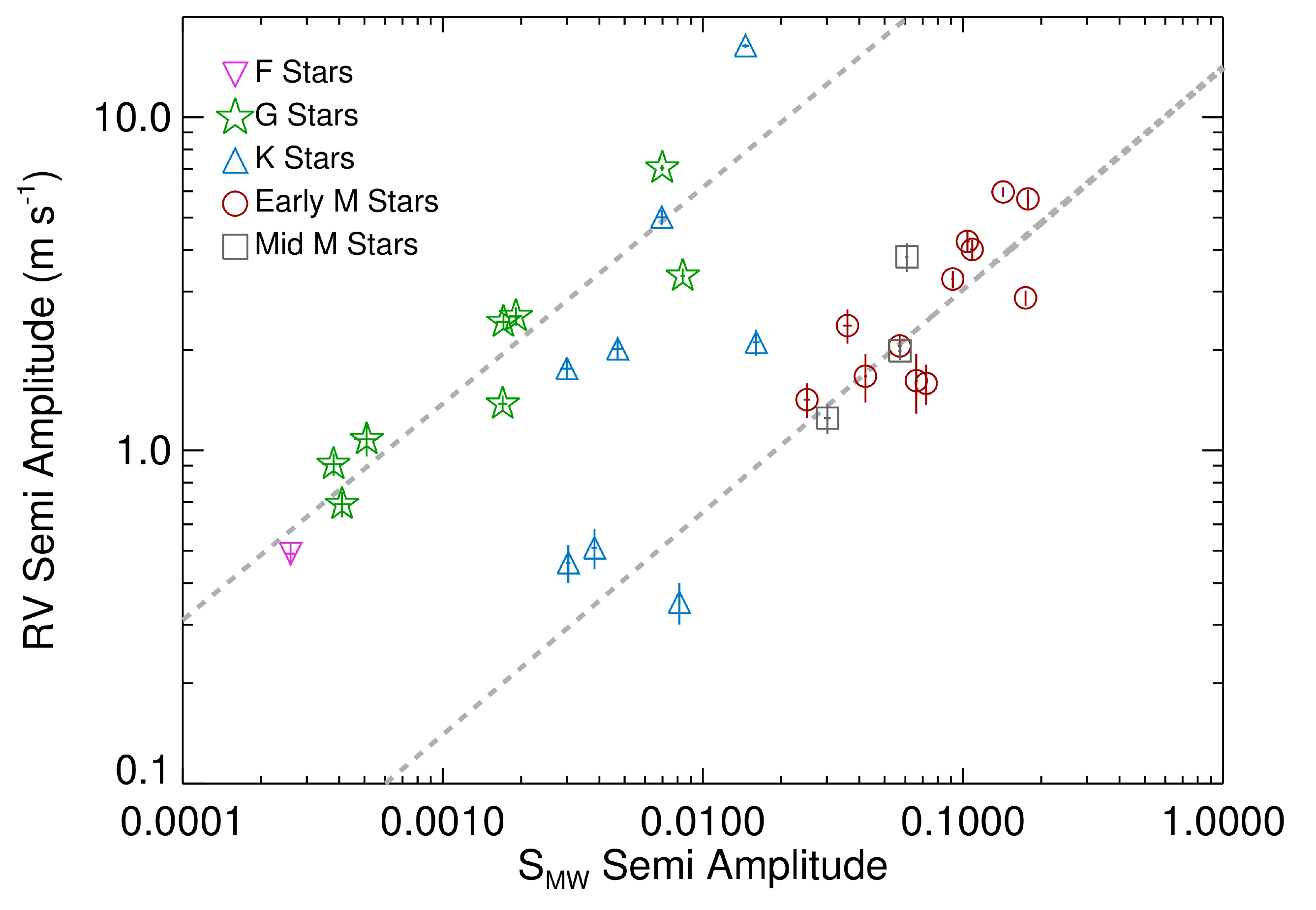}
\caption{Semi-amplitude of the rotation induced signal versus the semi-amplitude of the rotation signal in the S$_{MW}$ index. There are two distinct populations, with a break at K2.5-type stars.}
\label{amp_s_amp}
\end{figure}

In addition, we have studied the relationship between the semi-amplitude of the radial velocity signals and the semi-amplitude of their related activity signals. We found that the amplitude in radial velocity correlates with the amplitude in the S$_{MW}$ index signals. Figure~\ref{amp_s_amp} shows how the semi-amplitude of the radial velocity signals compare with the semi-amplitude of the S$_{MW}$ index signals. Even with our limited sample two distinct populations seem quite clear. F to early K stars and mid-K stars to mid-M dwarfs. The jump happens around spectral type K2.5, which corresponds to an effective temperature of 4800 K. This is the surface temperature when the relative abundance of Ca II drops dramatically until being marginal at $\sim$4000 K. FG and early K in one side, and late-K and M-type stars on the other are fit independently using the following equation:
\begin{equation}
   K_{V_{r}}= a \cdot K_{S_{MW}} + b
   \label{eq_rv_ind_2}
\end{equation}

Table~\ref{fit_parameters_2} shows the measured values for the parameters in equation~\ref{eq_rv_ind_2}.

\begin {table}
\begin{center}
\caption {Parameters for equation~\ref{eq_rv_ind_2}}
    \begin{tabular}{ l  c  c } \hline
Dataset & a & b  \\ \hline
FG and early K-type stars & 0.72 $\pm$ 0.02 & 2.36 $\pm$ 0.04\\
Late K and M-type stars & 0.89 $\pm$ 0.02 & 1.41 $\pm$ 0.02\\ \hline
\label{fit_parameters_2}
\end{tabular}  
\end{center}
\end {table}

\begin{figure}
\includegraphics[width=\linewidth]{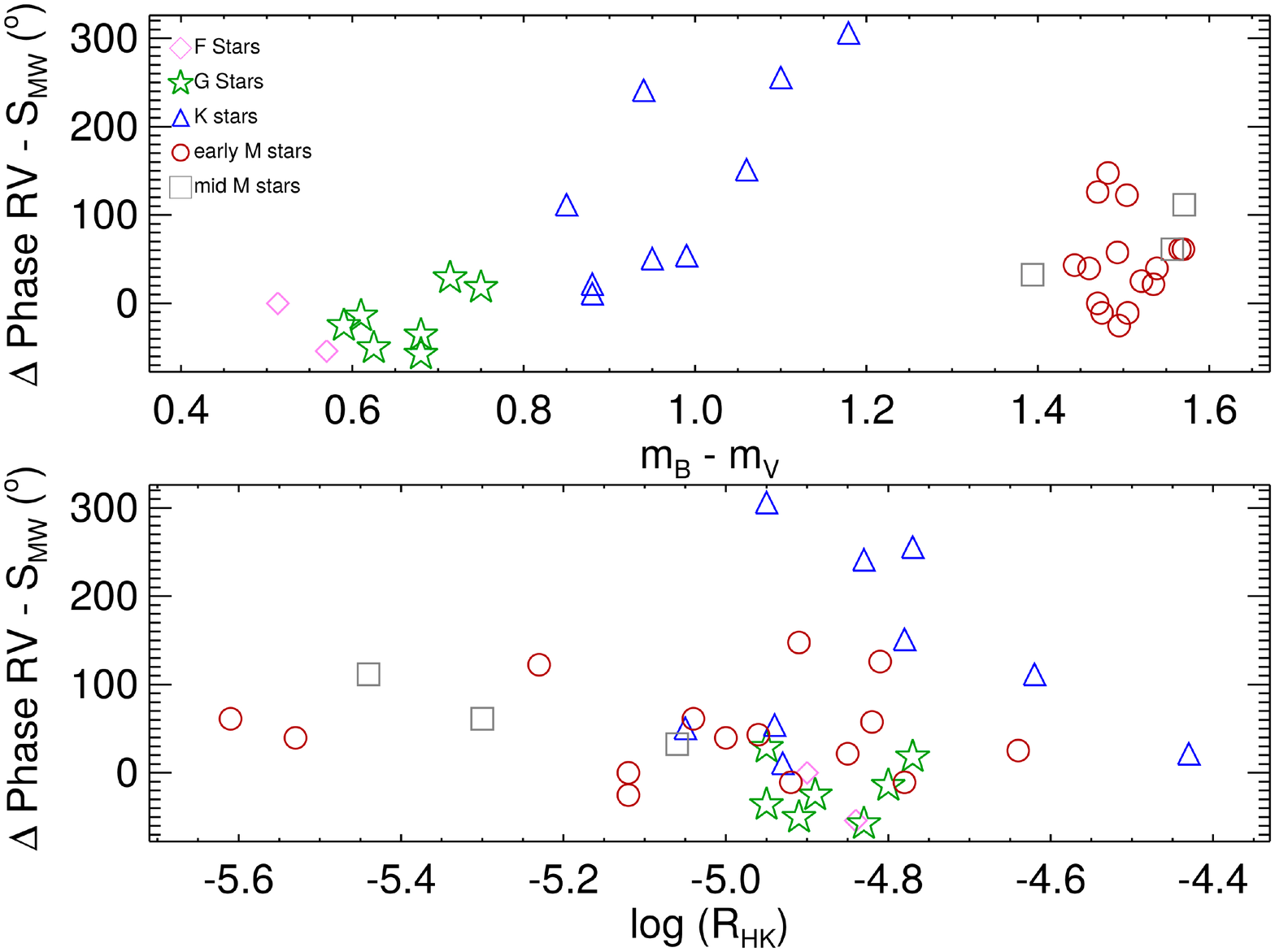}
\caption{Phase shift between the RV and the S$_{MW}$ signals against the color B-V (top panel) and the mean $\log_{10}(R'_\textrm{HK})$ (bottom panel).}
\label{phase_diff}
\end{figure}

Finally, we study the phase difference between the induced RV periodic signals and the periodic modulation in the activity indicators. \citet{Bonfils2007} and \citet{Santos2014} found that these signals are not necessarily in phase. In Figure~\ref{phase_diff} we show the phase difference between the RV signal and the S$_{MW}$ signal against the color of the stars and the mean activity level. We restrict this analysis to those stars where the period estimates of different indicator were consistent within error bars. We see an apparent evolution of the phase shift with the radial velocity signal lagging behind the S$_{MW}$ signal towards redder stars. Starting in G-type stars, which show small dispersion close to shift zero, and gradually increasing for K-type which reach a phase shift close to 360 degrees (zero again) for mid/late K-type stars, a trend which is seems to be continued for M-type stars. We do not find any correlation between the phase shift and the chromospheric activity level of the stars. Further investigation on a larger sample will be needed to confirm this behaviour.

Then we study the phase shift between the RV signal and the H$_{\alpha}$ index signal finding no clear relationship.

We also study the phase shift between the H$_{\alpha}$ index and the S$_{MW}$ index and find a group of the stars (HD 30495, HD 59468, Corot-7, HD 209100, GJ 676A, GJ 536, GJ 667C, GJ 674 and GJ 526) show a difference in phase close to phase 180 degrees, while the rest of the stars show a difference in phase close to zero. We do not see a clear correlation between the phase shifts and the average level of activity of each of these two groups.

The distinct behaviour of the different spectral types seen in Fig.~\ref{amp_rhk} and Fig.~\ref{amp_s_amp}, along with phase shift between the RV signal and the S$_{MW}$ signal (Figure~\ref{phase_diff}), give a clue about the nature of the dominant surface features causing the induced RV variations. The relationship between the Ca II H\&K emission and the effective change in radial velocity varies for the different spectral types. Ca II H\&K emission is generated mainly in the stellar plages \citep{ShineLinsky1974}, which are the main source of stellar induced RV variations in slowly rotating solar type stars \citep{Lockwood2007, Shapiro2014, Dumusque2014}. In this scenario, and assuming a high convection level - as in the case of the Sun - it is expected the RV signal and the activity signal to be more or less in phase, as we are seeing in our G-type stars. In cooler stars it seems that spots become more and more important. A radial velocity variation cased solely by spots would be expected to show a 90 degrees shift between the RV signal and the activity index signal. The gradual change in the phase might tell us that the equilibrium between the contributions of both types of activity regions gradually changes towards smaller stars, with more complex contributions causing a wide variety of phase shifts between the two signals.

\section{Conclusions}

Using data for 55 late-type stars (from late-F to mid-M) we have analysed the radial velocity time-series searching for periodic signals that match the stellar rotation periods. For 37 stars we have clearly found the induced RV signal by the stellar rotation or one of its harmonics, and measured its semi-amplitude.

Our study supports previously reported doubts on the keplerian origin of the periodic RV signals attributed to planets HD 40307 e, GJ 676 A e, GJ 667C d and GJ 832 c, which can also be explained as rotation-induced activity signals.

We have investigated the correlation between the level of chromospheric emission, represented by the $\log_{10}(R'_\textrm{HK})$ index, and the measured semi-amplitude, and obtained a specific linear relationship between this index and the logarithm of the amplitude for each spectral type. 

We have also investigated the correlation between the amplitude of the radial velocity signals and the amplitude in the activity signals, and we find two different correlations for G to mid-K stars and late-K to mid M dwarfs.

We have studied the phase shift between the period radial velocity induced signal caused by stellar rotation and the period signal from the S$_{MW}$ index. We find an apparent evolution of the phase shift and the color B-V. We do not find a correlation between the phase shift and the activity level.

The systematic measurement and characterisation of stellar activity induced RV signals is a necessary step for a reliable identification of the RV signals produced by terrestrial planets at short orbits. For the same activity level, M-type stars show larger activity-induced RV signals than G and K-type stars. Because of their lower stellar mass, both very low activity M dwarfs and late K dwarfs offer a very good opportunity for the detection of terrestrial planets.

\section*{Acknowledgements}

This work has been financed by the Spanish Ministry project MINECO AYA2014-56359-P. J.I.G.H. acknowledges financial support from the Spanish MINECO under the 2013 Ramón y Cajal program MINECO RYC-2013-14875. This work is based on data obtained from the HARPS public database at the European Southern Observatory (ESO). This research has made extensive use of the SIMBAD database, operated at CDS, Strasbourg, France and NASAs Astrophysics Data System. We are grateful to all the observers of the following ESO projects, whose data we are using 60.A-9036, 072.C-0096, 073.C-0784, 073.D-0038, 073.D-0578, 074.C-0012, 074.C-0364, 074.D-0131, 075.D-0194, 076.C-0878, 076.D-0130, 076.C-0155, 077.C-0364, 077.C-0530, 078.C-0044, 078.C-0833, 078.D-0071, 079.C-0681, 079.C-0927, 079.D-0075, 080.D-0086, 081.C-0148, 081.D-0065, 082.C-0212, 082.C-0308, 082.C-0315, 082.C-0718, 083.C-1001, 083.D-0040, 084.C-0229, 085.C-0063, 085.C-0019, 085.C-0318,
086.C-0230, 086.C-0284, 087.C-0368, 087.C-0831, 087.C-0990, 088.C-0011, 088.C-0323, 088.C-0353, 088.C-0662, 089.C-0050, 089.C-0006, 090.C-0421, 089.C-0497, 089.C-0732, 090.C-0849, 091.C-0034, 091.C-0866, 091.C-0936,  091.D-0469, 180.C-0886, 183.C-0437, 183.C-0972, 188.C-0265, 190.C-0027, 191.C-0505, 191.C-0873, 282.C-5036.

\bibliography{RHK_ref}

\label{lastpage}

\end{document}